\documentclass[aps,floatfix,preprint,nofootinbib, preprintnumbers, superscriptaddress,showpacs]{revtex4}
\usepackage[parfill]{parskip}    
\usepackage{graphicx}
\usepackage{amssymb}
\usepackage{epstopdf} 
\usepackage{verbatim}
\usepackage{url}
\newcommand{\be}{\begin{equation}}
\newcommand{\bea}{\begin{eqnarray}}
\newcommand{\beq}[1]{\begin{equation}\label{#1}}

\newcommand{\ee}{\end{equation}}
\newcommand{\eea}{\end{eqnarray}}
\newcommand{\eeq}{\end{equation}}
\newcommand{\lsim}{\!\mathrel{\hbox{\rlap{\lower.55ex \hbox{$\sim$}} \kern-.34em \raise.4ex \hbox{$<$}}}}
\newcommand{\gsim}{\!\mathrel{\hbox{\rlap{\lower.55ex \hbox{$\sim$}} \kern-.34em \raise.4ex \hbox{$>$}}}}
\newcommand{\mr}[1]{\mathrm{#1}}

\begin{document}

\setlength{\baselineskip}{0.22in}

\preprint{MCTP-11-33} 
\preprint{ SLAC-PUB-14584}

\title{Singlet--Doublet Dark Matter}
\author{Timothy Cohen}
\affiliation{
SLAC National Accelerator Laboratory,
2575 Sand Hill Rd, Menlo Park, CA 94025}
\affiliation{Michigan Center for Theoretical Physics (MCTP) \\
Department of Physics, University of Michigan, Ann Arbor, MI
48109}
\author{John Kearney and Aaron Pierce}
\vspace{0.2cm}
\affiliation{Michigan Center for Theoretical Physics (MCTP) \\
Department of Physics, University of Michigan, Ann Arbor, MI
48109}
\author{David Tucker-Smith}
\vspace{0.2cm}
\affiliation{Williams College Department of Physics,
Williams College, Williamstown, MA 01267}

\date{\today}

\begin{abstract}
In light of recent data from direct detection experiments and the Large Hadron Collider, we explore models of dark matter in which an $SU(2)_{L}$ doublet is mixed with a Standard Model singlet.  We impose a thermal history. If the new particles are fermions, this model is already constrained 
due to null results from  XENON100.  
We comment on remaining regions of parameter space and assess prospects for future discovery.  We do the same for the model where the new particles are scalars, which at present is less constrained.  Much of the remaining parameter space 
for both models will be probed by the next generation of direct detection experiments.  
For the fermion model, DeepCore may also play an important role.
\end{abstract}
\maketitle

\section{Introduction}
A weakly interacting massive particle (WIMP) remains an attractive candidate to explain dark matter.
But what exactly is meant by  ``weakly?" 
Often, all that is implied is that annihilation cross sections are parametrically suppressed by the weak mass scale, $\sigma_\mr{ann} \sim m_W^{-2}$; the precise mechanism of annihilation may or may not involve the bosons of the electroweak theory. 
As an example consider supersymmetry, where
 annihilations may be mediated by particles of
the supersymmetric sector.   

In this paper we address the following question: does a 
strictly weakly
interacting particle, i.e., one whose annihilation is controlled by the $W$, $Z$ and Higgs bosons, remain an attractive dark matter candidate?  Such a dark matter candidate would not require the introduction of new mediators, and would thus provide a well-motivated, economical scenario.  A particle possessing full-strength interactions with the $Z$ boson,  e.g. a heavy Dirac neutrino, would have a direct detection cross section many orders of magnitude in excess of present limits.  A simple remedy is to mix a sterile state with 
this active state.  This mixing yields
two effects:  it reduces the size of the coupling to the gauge bosons and, in the case of fermions, can transform
the dark matter from a Dirac particle into a Majorana particle.  Together, these variations
enable the dark matter to have both an annihilation cross section consistent with a thermal history and a direct detection cross section that is not yet excluded.  In supersymmetry, the bino may play the role of this sterile state, and can be mixed with the Higgsinos
to achieve a well-tempered neutralino, a possibility emphasized in \cite{ArkaniHamed:2006mb}.  For a different approach to strictly weakly interacting Dark Matter, see \cite{Cirelli:2005uq}.

Here, we do not confine ourselves to 
supersymmetric models, but instead explore more generically the consequences of mixing a Standard Model singlet with an active particle.
The particular case where the 
charged
state has the quantum numbers of 
a doublet is worthy of special attention.  In this case, the mixing can naturally be provided by a renormalizable coupling to the Higgs field.  This fermionic singlet--doublet model has been previously explored in the literature  \cite{ArkaniHamed:2005yv,Mahbubani:2005pt,D'Eramo:2007ga,Enberg:2007rp}, and serves to inform us about the viability of 
strictly weakly
interacting dark matter in light of recent improvement in direct detection bounds and the negative searches for the Higgs boson at the Large Hadron Collider (LHC).  We also consider the scalar analog of this model, in which a scalar doublet is mixed with a real scalar singlet\cite{Kadastik:2009dj,Kadastik:2009cu}.

After imposing a thermal history, much of the parameter space for the fermionic model has been excluded.
To avoid tension with direct detection bounds,  we find one of the following exceptional cases must apply:
\begin{enumerate}
\item{The dark matter mass could be close to half the mass of either the Higgs or $Z$ boson.}
\item{Masses in the dark matter sector could be arranged such that co-annihilation is important.}
\item{The couplings to the Higgs boson could be small.  This does not necessarily imply that the couplings that induce the mixing are small, as there is room for non-trivial cancellations.}
\item{The Higgs boson could
be heavy.  This can be made consistent with precision electroweak constraints without the need for any additional physics, since
this model can give a large positive contribution to the $T$ parameter in a straight-forward way \cite{D'Eramo:2007ga,Enberg:2007rp}.}
\end{enumerate}
We explore these possibilities in detail in Sec.~\ref{sec:fermion}.  Recent data from the LHC have had an impact on the fourth possibility.  ATLAS \cite{leptonphotonATLAS2011} and CMS \cite{leptonphotonCMS2011} have greatly constrained the range of allowed values for the Higgs boson mass, $m_h$.  A naive combination of the results from these experiments disfavors Higgs boson masses in the range $150 \text{ GeV} \lesssim m_h \lesssim 450 \text{ GeV}$.  Consequently, to avoid direct detection bounds by making the Higgs boson heavy, i.e., heavier than $\sim 150\mbox{ GeV}$,  now requires a significant increase in the Higgs boson mass.  Motivated by these findings, we mainly consider two
scenarios: a light Higgs boson ($m_h = 140 \text{ GeV}$), and a heavy Higgs boson ($m_h = 500 \text{ GeV}$).  We also comment on an intermediate case ($m_h = 200 \text{ GeV}$) in which the dark sector could conceivably contribute significantly to the invisible width of the Higgs boson such that the recent experimental bounds are evaded.
Both spin-independent and spin-dependent direct detection searches will be important future probes of this model.

The physics of the scalar model can be quite different.  For instance,  because of the possible presence of a singlet--Higgs boson mixed quartic, no mixing is necessary to achieve a dark matter-Higgs boson coupling. While at present this scalar model
is less constrained, spin-independent direct detection experiments will probe much of its parameter space in the near future.  We examine this model in Sec.~\ref{sec:scalar}.

\section{The singlet doublet fermion model}\label{sec:fermion}

We consider an extension of the Standard Model consisting of a gauge singlet 
fermion 
and a pair of
fermionic
 electroweak doublets.  The doublets have a vector-like mass term, and the neutral components of the doublets mix with the gauge singlet through renormalizable couplings to the Higgs boson.  
These fields are odd
under a $\mathbb{Z}_2$ symmetry,  ensuring the stability of the lightest state.  
We denote the singlet as $S$ and the doublets as $D$ and $D^c$:
\begin{eqnarray}
D = \left(\begin{array}{c} \nu \\ E \end{array} \right) & \qquad & D^c = \left(\begin{array}{c} -E^c \\ \nu^c \end{array}\right),
\end{eqnarray}
with hypercharges $-\frac{1}{2}$ and $+\frac{1}{2}$ respectively,
implying that the $\nu$ states are electrically neutral.  
Mass terms and interactions for this model are given by:
\begin{equation}
\Delta \mathcal{L} = - \lambda D H S - \lambda' D^c \tilde{H} S - M_D D D^c - \frac{1}{2} M_S S^2 + \mbox{ h.c.},
\end{equation}
where $SU(2)$ doublets are contracted with the Levi-Civita symbol $\epsilon^{ij}$ and $\tilde{H} \equiv i \sigma_2 H^\ast$.  Field re-definitions leave one physical phase for the set of parameters $\{M_S, M_D, \lambda, \lambda'\}$.  For simplicity we take them to be real.  
Discussions of the consequences of introducing a non-zero phase may be found in \cite{Mahbubani:2005pt,D'Eramo:2007ga}.   As alluded to in the introduction, in addition to being an interesting candidate for dark matter in its own right,  this model is similar to neutralino dark matter in the MSSM (or Split Supersymmetry), in which the sterile Bino mixes with the electroweak doublet Higgsinos (in the limit where the Wino decouples, $M_2 \rightarrow \infty$).  Consequently, it provides a laboratory where one can potentially gain insight into the physics of MSSM dark matter.\footnote{In fact, \cite{Fayet:1974fj}, where a singlet-doublet model was considered (but without a majorana mass for $S$), was an important historical step on the road towards supersymmetric electroweak theories  \cite{Fayet:1974pd}.}

Expanding 
the Higgs field 
around
its vacuum expectation value, $v=246 \text{ GeV}$, we can write the neutral mass terms
in the basis $\psi^0 = (S, \nu, \nu^c)$ as:
\begin{equation}
\Delta \mathcal{L} \supset - \frac{1}{2} (\psi^0)^T \mathcal{M} \psi^0 + \text{h.c.} = - \frac{1}{2} (\psi^0)^T \left(\begin{array}{ccc} M_S & \frac{\lambda}{\sqrt{2}} v & \frac{\lambda'}{\sqrt{2}} v \\ \frac{\lambda}{\sqrt{2}} v & 0 & M_D \\ \frac{\lambda'}{\sqrt{2}} v & M_D & 0 \end{array} \right)\psi^0 + \text{h.c.}
\end{equation}
It can also be instructive to write this in terms of the rotated basis $\psi^0_r = (S,\frac{\nu^c+\nu}{\sqrt{2}},\frac{\nu^c-\nu}{\sqrt{2}})$:
\begin{equation}
\Delta \mathcal{L} \supset - \frac{1}{2} (\psi^0_r)^T \left(\begin{array}{ccc} M_S & \frac{\lambda_+}{2} v & \frac{\lambda_-}{2} v \\ \frac{\lambda_+}{2} v & M_D & 0 \\ \frac{\lambda_-}{2} v & 0 & - M_D \end{array}\right) \psi^0_r + \text{h.c.}
\label{eq:rotatedmassmatrix}
\end{equation}
where $\lambda_\pm = \lambda' \pm \lambda$.  The three physical mass eigenstates for the neutral particles are a linear combination of singlet and doublet states:\footnote{We agree with the expressions for the masses and mixing angles given in \cite{Enberg:2007rp} with the caveat that the third mass eigenvalue given in their Eq.~(A.1) corresponds to the mass of the lightest particle, and the first to the mass of the heaviest.}
\begin{equation}
\nu_i = \vartheta_i S + \alpha_i \nu + \beta_i \nu^c, \; \; \; (i= 1,2,3).
\label{eqn:composition}
\end{equation}
We let $\nu_1$ denote the lightest (Majorana) neutral state --- this is our dark matter candidate.  The spectrum also contains a Dirac fermion $\psi_E$ composed of the fields $E$ and $E^c$ with mass $M_D$.

\begin{figure}
\includegraphics[width=\textwidth]{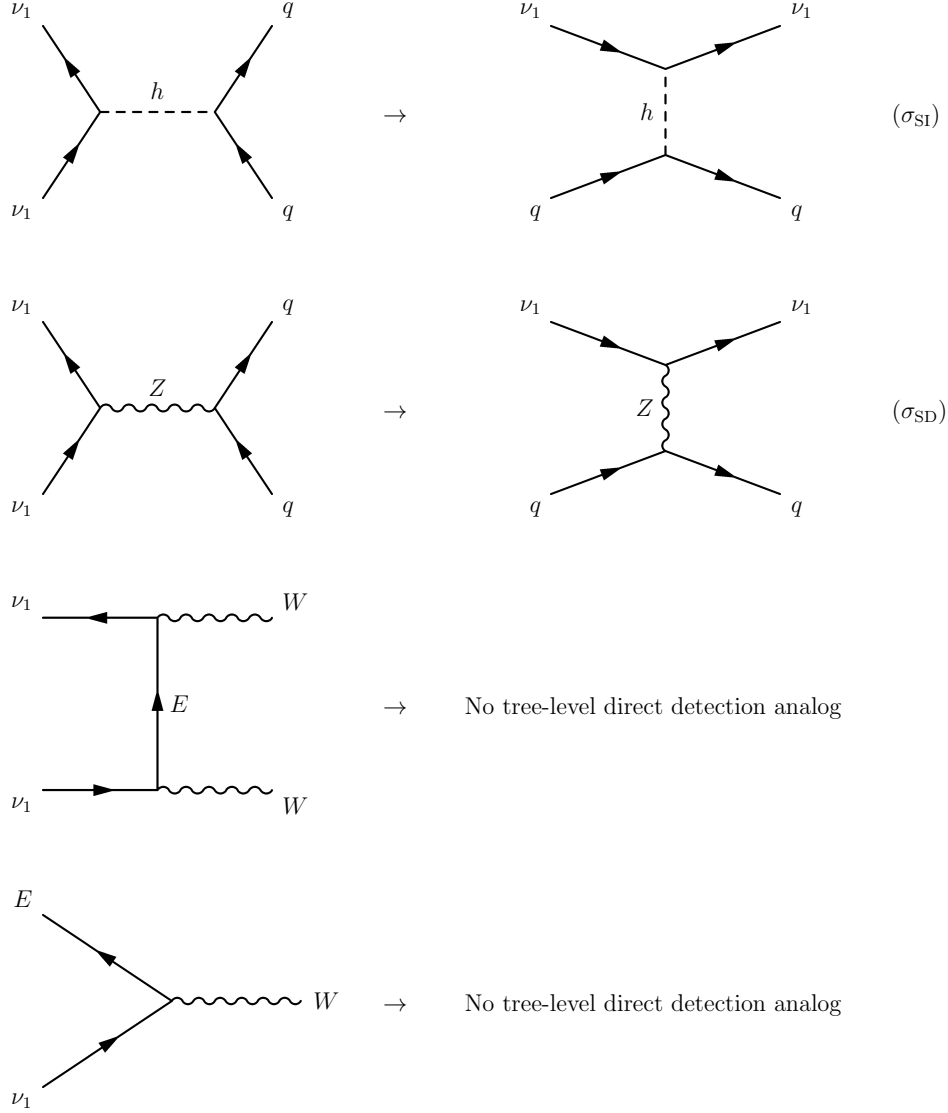}
\vspace{-2in}
\caption{\label{fig:FeynSym}Relevant diagrams for annihilation and corresponding direct detection diagrams, where applicable.  Achieving sufficient dark matter annihilation in the early universe in order to obtain the measured
relic density requires at least one of these diagrams to be significant.  In the case of $s$-channel Higgs or $Z$ boson exchange, this may imply correspondingly large $\sigma_{\text{SI}}$ or $\sigma_{\text{SD}}$ respectively.  In the case of $t$-channel annihilation or co-annihilation, there is not a clear direct detection analog, but the processes will be related through couplings and mixing angles.}
\end{figure}

As a linear combination of singlet and doublet states, $\nu_1$  generically has a coupling to the Higgs boson and a coupling to the $Z$.
These couplings can provide channels for dark matter annihilation in the early universe through $s$-channel Higgs and $Z$ boson exchange.
If the $\nu_1 \nu_1 h$ coupling is considerable, this coupling may also yield a large spin-independent cross section.
Rotating the Feynman diagram for annihilation of dark matter to quarks via an $s$-channel Higgs boson produces a diagram that contributes to spin-independent direct detection, as illustrated in Fig.~\ref{fig:FeynSym}.  Similarly, a large $\nu_1 \nu_1 Z$ coupling may yield a 
large spin-dependent cross section.  This is a salient feature of strictly WIMP dark matter --- generically, the mediators responsible for annihilation 
($h$ and $Z$, in particular) also couple to protons, which can result in observable
direct detection signals.  However, there do exist additional processes by which the dark matter can annihilate in the early universe, including annihilation directly to gauge bosons via $t$-channel exchange of various beyond the Standard Model particles (for $m_{\nu_1} > m_W$), and co-annihilation \cite{Griest:1990kh}.  These processes are also illustrated in Fig.~\ref{fig:FeynSym}, and unlike the $s$-channel processes have no tree-level direct detection analog.  That said, the couplings involved depend on the mixing angles, so there can still be non-trivial correlations between dark matter annihilation in the early universe and direct detection cross sections.

Recent data from direct detection experiments, notably XENON100 \cite{Aprile:2011hi} and SIMPLE \cite{Felizardo:2011uw}, have substantially improved the sensitivity to both spin-independent and -dependent scattering with no evidence for the detection of dark matter.   Although  DAMA \cite{Bernabei:2008yi}, CoGeNT \cite{Aalseth:2011wp}, and CRESST \cite{Angloher:2011uu} have reported possible evidence for dark matter scattering, this interpretation seems to be in serious tension with null results from XENON100, CDMS and EDELWEISS \cite{Ahmed:2011gh}, and other direct detection experiments, and a  coherent explanation for  these possible signals  is lacking at present.  It is conceivable that a consistent picture may one day emerge, but
in this paper we operate under the assumption that existing data do not indicate signals, and dark matter detection cross sections should lie beneath current bounds.  

\subsection{Relic density and cross section calculations}\label{sec:calculate}

Given the above discussion it is interesting to ask 
whether this simple WIMP model always has large direct detection signals, or whether it is possible to have highly
suppressed spin-independent cross sections $\sigma_{\mr{SI}}$ and/or spin-dependent cross sections $\sigma_{\mr{SD}}$\footnote{Throughout this paper, $\sigma_\mr{SI}$ is strictly the cross section off of the proton, but for the class of models considered the spin-independent cross sections off the proton and neutron are equal to an excellent approximation}.  To address this and related questions
we calculate relic densities and direct detection cross sections in \texttt{micrOMEGAs}2.4 \cite{Belanger:2010gh}, using our implementation of the relevant models. 
\texttt{micrOMEGAs} employs the following values for the scalar nuclear matrix elements:
\begin{eqnarray*}
f_{Tu}^{(p)} = 0.023 \quad f_{Td}^{(p)} = 0.033 \quad f_{Ts}^{(p)} = 0.259 \\
f_{Tu}^{(n)} = 0.018 \quad f_{Td}^{(n)} = 0.042 \quad f_{Ts}^{(n)} = 0.259&,
\end{eqnarray*}
although recent lattice measurements suggest that smaller values may be more accurate, which would weaken direct detection bounds \cite{Giedt:2009mr}.  The above choices correspond to an effective Higgs boson--proton coupling of $f$ =0.467, whereas the lattice evaluation corresponds to a $f=0.30 \pm 0.015$ \cite{Giedt:2009mr}.  The difference is a decrease of the quoted spin-independent cross sections by roughly a factor of 2.5.

It is worth mentioning two approximations employed by \texttt{micrOMEGAs}.  First, \texttt{micrOMEGAs} does not include loops effects or the (velocity suppressed) contribution to the spin-independent cross section due to $Z$ exchange (the $(\bar{\psi}_{\nu_1} \gamma_\mu \gamma^5 \psi_{\nu_1}) (\bar{q} \gamma^\mu q)$ effective operator).  While these contributions are generally sub-dominant to those due to Higgs boson exchange, if the $\nu_1 \nu_1 h$ coupling were to be suppressed, these 
effects would play a significant role in determining 
$\sigma_{\text{SI}}$.  
Since the spin-independent cross sections produced by such effects tend be well below the current bounds \cite{Essig:2007az,Hisano:2011cs,Cirelli:2005uq}, we neglect these effects throughout our paper.  Rather, spin-independent cross sections 
$\lesssim 10^{-10} \text{ pb}$
 should be taken as illustrative of the very small direct detection cross sections at these points, and not as precise values.  A similar caveat holds for tiny spin-dependent cross sections. Second, it should be noted that \texttt{micrOMEGAs} accounts only for two-to-two scattering when computing the relic abundance.  Three-body processes can be relevant near the opening of a new channel, see e.g.~\cite{Yaguna:2010hn}.  For instance, 
as $m_{\nu_1}\rightarrow m_W$,
the $\nu_1 \nu_1 \rightarrow W W^\ast$ annihilation channel can become particularly relevant, but will be neglected in our calculations.  Similarly, 
as $m_{\nu_1} \rightarrow m_t$,
the $\nu_1 \nu_1 \rightarrow t t^\ast$ final state can become relevant.  This is especially important for dark matter that annihilates through an $s$-channel $Z$ boson, as the $\nu_1 \nu_1 \rightarrow Z \rightarrow t t$ process does not suffer from  $p$-wave suppression.

\subsection{Suppression of $\sigma_{\text{SI}}$ and $\sigma_{\text{SD}}$}\label{sec:suppressed}

For certain values of the parameters, it is indeed possible to cancel the tree-level coupling of the dark matter to the Higgs or $Z$ bosons, thereby realizing suppressed $\sigma_{\text{SI}}$ and $\sigma_{\text{SD}}$ respectively.  The case of the $Z$ is straightforward: the $\nu_1 \nu_1 Z$ coupling goes as $(\alpha_1^2 - \beta_1^2)$ in the notation of Eq.~(\ref{eqn:composition}).  Thus, whenever $\nu_1$ contains approximately equal amounts of $\nu$ and $\nu^c$ the coupling to the $Z$ boson will be small.  This occurs for either $\lambda_+ =0$ or $\lambda_- = 0$.  From Eq.~(\ref{eq:rotatedmassmatrix}), we see that in either
case mixing occurs between the $S$ and only one of the rotated doublet states, $\frac{\nu^c \pm \nu}{\sqrt{2}}$.  Consequently all neutral states mix with either $\frac{\nu^c + \nu}{\sqrt{2}}$ or $\frac{\nu^c - \nu}{\sqrt{2}}$, 
 meaning they will contain equal amounts of $\nu$ and $\nu^c$, and thus the $\nu_1 \nu_1 Z$ coupling will vanish.
$\lambda_\pm = 0 \Rightarrow \lambda' = \pm \lambda$ corresponds to the maintenance of a custodial $SU(2)$ symmetry in the new sector.  

We now derive the condition for eliminating the coupling between the Higgs boson and $\nu_1$.  For $M_S < M_D$, the mass of the lightest neutral particle can be written as:
\begin{equation}
m_{\nu_1} = M_S + v\, f(M_S, M_D, \lambda\, v, \lambda' v).
\end{equation}
By gauge invariance, the $\nu_1 \nu_1 h$ coupling is also proportional to $f$.  Thus, a choice of parameters parameters that satisfies $m_{\nu_1} = M_S$ for $M_S < M_D$ also eliminates the coupling to the Higgs boson.  The following relationship, derived from the characteristic mass eigenvalue equation,  cancels the $\nu_1 \nu_1 h$ coupling:
\begin{equation}
\lambda'_{\text{crit}} = - \lambda \frac{M_S}{M_D} \left(1 \pm \sqrt{1 - \left(\frac{M_S}{M_D}\right)^2}\right)^{-1}.
\end{equation}
Note, for $M_S < M_D$, it is not possible to simultaneously satisfy this condition and one of the conditions $\lambda_+=0$ or $\lambda_-=0$.
In other words, it is impossible in this case to simultaneously cancel the $\nu_1 \nu_1 h$ and $\nu_1 \nu_1 Z$ couplings.  

\begin{figure}
\begin{center}
\includegraphics[width=1.0\textwidth]{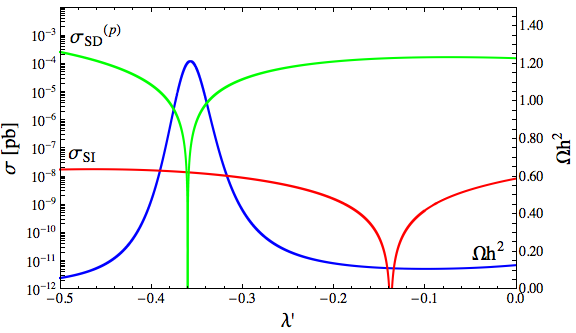}
\end{center}
\caption{\label{fig:Troughs}An example of the suppression of $\sigma_{\text{SI}}$ and $\sigma_{\text{SD}}$ as a function of $\lambda'$ for $M_S = 200 \text{ GeV}$, $M_D = 300 \text{ GeV}$ and $\lambda = 0.36$.  The critical value for $\nu_1 \nu_1 h$ cancellation is $\lambda' = -0.138$, and for $\nu_1 \nu_1 Z$ cancellation is $\lambda' = \pm 0.36$.  The lines shown are $\sigma_\mr{SD}^{(p)}$ [green], $\sigma_\mr{SI}$ [red] and $\Omega h^2$ [blue]. }
\end{figure}

An example of these cancellations for $M_S < M_D$ is shown in Fig.~\ref{fig:Troughs}. 
There, we fix $M_S$, $M_D$, and $\lambda$, and vary $\lambda^{\prime}$. With $M_S = 200 \text{ GeV}$, $M_D = 300 \text{ GeV}$ and $\lambda = 0.36$, for most values of $\lambda^{\prime}$ the relic density is set by annihilation through an $s$-channel $Z$.  Consequently, for $\lambda' \approx -0.36 = -\lambda$ (where the $\nu_1 \nu_1 Z$ coupling cancels) 
the annihilation cross section decreases and there is a dramatic increase in the relic density.  
Meanwhile, aside from this special point, $s$-channel Higgs boson exchange does not contribute significantly to the dark matter annihilation.  Correspondingly, at the point $\lambda' \approx -0.138 = \lambda'_{\text{crit}}$ where the $\nu_1 \nu_1 h$ coupling vanishes, the relic density is essentially unaffected.
Since $\sigma_\mr{SI} \sim (\lambda^{\prime} - \lambda^{\prime}_{\text{crit}})^2$, even a 10\% ``accident" where $\lambda^{\prime}$ takes on values close to this critical value can have important implications for spin-independent direct detection.

For the alternative case where $M_D < M_S$, the analogous analysis reveals the condition for $\nu_1 \nu_1 h$ cancellation to be $\lambda'_{\text{crit}} = - \lambda \Rightarrow \lambda_+ = 0$ ($m_{\nu_1} = M_D$).  The resultant WIMP is $\nu_1 = \frac{1}{\sqrt{2}} \left(\nu^c + \nu\right)$, and has suppressed coupling to both the Higgs and $Z$ boson.  However, the dark matter particle retains a full-strength coupling to the charged dark sector fermion and the $W$ boson.  Because the $E$ fermion also has mass $M_D$, there is significant contribution to dark matter annihilation from co-annihilation with the charged state.\footnote{Note that, in fact, the $E$ will be slightly heavier than the WIMP due to Coulombic radiative corrections.}  As this coupling strength is fixed, to achieve the correct relic density the value of $M_D$ is constrained to $M_D \gtrsim 1 \text{ TeV}$.  This situation is similar to the case of ``pure'' Higgsino dark matter in the MSSM, for which $M_D \sim 1.1 \text{ TeV}$ yields the correct value of $\Omega h^2$.  So, there is the possibility  that $m_{\nu_1} \gtrsim 1 \text{ TeV}$ with heavily suppressed spin-independent and spin-dependent cross sections.  For instance, we find that for $M_S = 2 \text{ TeV}$ and $\lambda = - \lambda' = 0.2$, the correct relic density is achieved for $M_D = 1.1 \text{ TeV}$.  For this point, $\sigma_{\text{SI}}$ and $\sigma_{\text{SD}}$ are heavily suppressed as the $\nu_1 \nu_1 h$ and $\nu_1 \nu_1 Z$ couplings are small, and $m_{\nu_2} - m_{\nu_1} \sim 1 \text{ GeV}$, sufficiently large to effectively prohibit direct detection via inelastic scattering.  Incidentally, in contrast to the MSSM, the freedom to choose the size of the $\lambda$ coupling allows a wider range of (all heavy) $M_{D}$ values.

In models that have built-in relations between $\lambda$ and $\lambda^\prime$, such as the MSSM, there is a question as to whether these cancellations are still possible. 
In the MSSM, we find cancellations and an appropriate relic density are indeed simultaneously realizable, but only for small values of $\tan \beta$.  In particular, the $\lambda_+ = 0$ condition just discussed is achieved for $\tan \beta = 1$ (it is impossible to achieve $\lambda_- = 0$ due to the relative signs between off-diagonal couplings in the MSSM), and for $M_1 < \mu$ (analogous to $M_S < M_D$) we find the cancellation of the dark matter coupling to the Higgs boson and the correct relic density only for values of $\tan \beta \lesssim 2$.  Thus, in the MSSM there is tension between suppressing direct detection cross sections and generating a sufficiently large Higgs boson mass.  Amusingly, we find for $M_1 < \mu, M_2$, the high degree of symmetry between the off-diagonal entries in the neutralino mass matrix results in the condition for canceling the dark matter-Higgs boson coupling being the identical for any  $M_2 > M_1$.

Returning now to the singlet-doublet model, for a small $\nu_1 \nu_1 h$ coupling (and $\sigma_{\text{SI}}$) a sizeable $\nu_1 \nu_1 Z$ coupling (and $\sigma_{\text{SD}}$) might still be required to achieve sufficient dark matter annihilation in the early universe, or vice-versa.  We now investigate the general size of the direct detection cross sections for a dark matter relic density of $0.1053 \leq \Omega h^2 \leq 0.1193$, a {$2 \sigma$} range determined by the combination of the seven-year Wilkinson Microwave Anisotropy Probe (WMAP) and other data on large scale structure \cite{Jarosik:2010iu}.  In what follows we will investigate the differences in the dark matter phenomenology associated with a light versus a heavy Higgs boson.  This will provide us with a sense of the likelihood of discovery of this particular model as direct detection experiments increase in sensitivity in the coming years, and of the fate of fermionic WIMP dark matter in general.

\subsection{Light Higgs boson $m_h = 140 \text{ GeV}$}

Some previous studies of this model have focused on the possibility of new dark states charged under $SU(2)_L$ generating a large contribution to the oblique $T$ parameter \cite{Enberg:2007rp}.  For a relatively light Higgs boson with $m_h = 140 \text{ GeV}$, such a large contribution is undesirable.  We require the contribution to the $T$ parameter from the dark sector lie in the range:
\begin{equation}
-0.07 \leq \Delta T \leq 0.21
\end{equation}
Exact expressions for $\Delta T$ can be found in \cite{D'Eramo:2007ga}.
As in \cite{Enberg:2007rp}, we neglect the new physics contributions to $S$ and $U$, which are significantly smaller than the contributions to $T$.  The range given above represents the shift in $\Delta T$ required by the new physics to ensure that the oblique parameter values for the model remain within the 68\% ellipse in the $(S,T)$ plane.\footnote{This ellipse is larger than the restrictive 39.35\% ellipse shown in \cite{Nakamura:2010zzi}.}

\begin{figure}
\begin{center}
\includegraphics[width=0.75\textwidth]{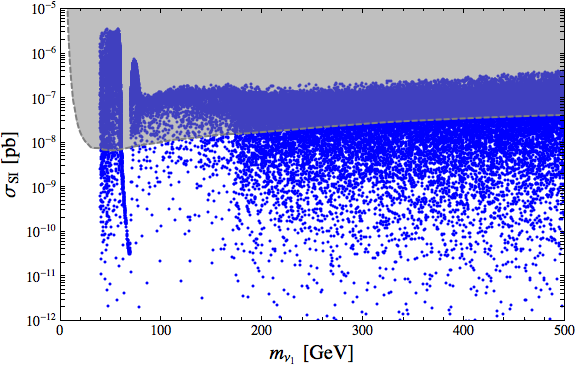}
\includegraphics[width=0.75\textwidth]{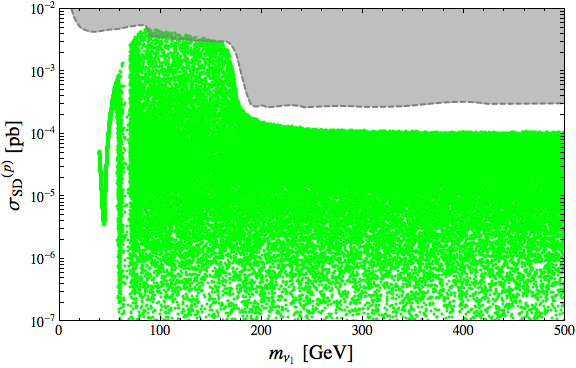}
\end{center}
\caption{\label{fig:SIDmassMH140}
Plots of spin-independent [top] and spin-dependent [bottom] cross sections against dark matter mass for $m_h = 140 \text{ GeV}$.  Points satisfy the thermal relic density constraint.  Shaded regions represent $\sigma_{\text{SI}}$ exclusion limits from XENON100 \cite{Aprile:2011hi} [top] and combined $\sigma_{\text{SD}}^{(p)}$ exclusion limits from SIMPLE, Super-K and IceCube (hard) \cite{Felizardo:2011uw,Desai:2004pq,Abbasi:2009uz} [bottom].  Exclusion curves assume a local dark matter density of $\rho = 0.3$ GeV/cm$^3$.} 
\end{figure}

We perform a random scan of the parameter space with $0 \text{ GeV} \leq M_S \leq 800 \text{ GeV}$, $80 \text{ GeV} \leq M_D \leq 2 \text{ TeV}$, $-2 \leq \lambda \leq 2$ and $0 \leq \lambda' \leq 2$.  We permit relatively large values of $\lambda$ and $\lambda'$ to avoid imposing any theory bias.  However, we note that restricting to smaller couplings $-1 \leq \lambda \leq 1$ and $0 \leq \lambda' \leq 1$ would not significantly alter the results.  In addition to requiring the relic density to be in the range $0.1053 \leq \Omega h^2 \leq 0.1193$, we require the mass of the dark matter to be $40 \text{ GeV} \leq m_{\nu_1} \leq 500 \text{ GeV}$.  Points with $m_{\nu_1}$ much less than 40 GeV would typically lead to an excessive contribution to the invisible width of the $Z$.   This contribution can be turned off by setting $\lambda' = \pm \lambda$.  However, doing so leaves Higgs boson exchange as the only annihilation process in the early universe, and for these small values of $m_{\nu_1}$ it turns out that Higgs boson exchange alone cannot yield a realistic relic density.

Plots of $\sigma_{\text{SI}}$ and $\sigma_{\text{SD}}^{(p)}$ against $m_{\nu_1}$ are shown in Fig.~\ref{fig:SIDmassMH140}, along with exclusion limits from XENON100 \cite{Aprile:2011hi} and SIMPLE/Super-K/IceCube \cite{Felizardo:2011uw,Desai:2004pq,Abbasi:2009uz}.  The exclusion curves shown assume a local dark matter density of $\rho = 0.3$ GeV/cm$^3$.  A recent evaluation suggests a somewhat higher density \cite{Salucci:2010qr}, which would give rise to proportionally stronger bounds.  It should be noted that the spin-dependent limits shown for $m_{\nu_1} \gtrsim m_W$ are model-dependent indirect detection limits, which assume certain dark matter annihilation channels in the Sun and Earth.  The limits shown for $m_W \lesssim m_{\nu_1} \lesssim m_t$ are taken directly from Super-K's paper \cite{Desai:2004pq}.  They assume dark matter annihilations to $\tau \tau$ (presumably neglecting neutrino oscillations).  For the points in Fig.~\ref {fig:SIDmassMH140} near these limits, the dark matter has sizable $\nu_1 \nu_1 Z$ coupling, and will exhibit dominant annihilation in the Sun and Earth via an $s$-channel $Z$ to $b b$, which dominates in the $v \rightarrow 0$ limit.   There is non-trivial annihilation to $\tau \tau$ as well, but annihilation to W boson pairs is tiny due to velocity suppression.  So, while the limits are representative, they are not precise. If $m_{\nu_1} \gtrsim m_t$, the hard limits shown from IceCube assume annihilation to $W W$.  In fact, in this region, the points nearest the limits will be once again be characterized by Dark Matter that annihilates predominantly via an $s$-channel $Z$, although in this case to $t t$, in the Sun and Earth.  The tops will decay to produce fairly hard $W$ bosons, so in this case the limits shown are representative of the actual model-dependent limits but the actual limits will be slightly weaker.

\begin{figure}
\begin{center}
\includegraphics[width=0.75\textwidth]{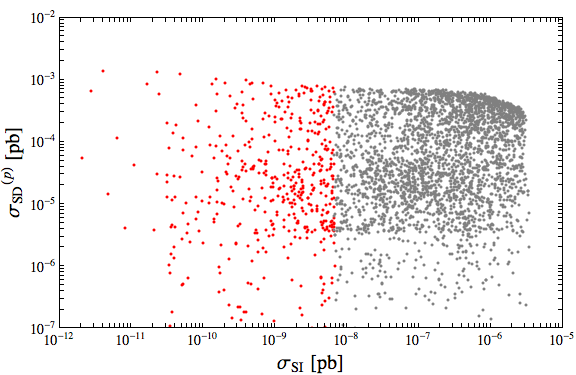}
\includegraphics[width=0.75\textwidth]{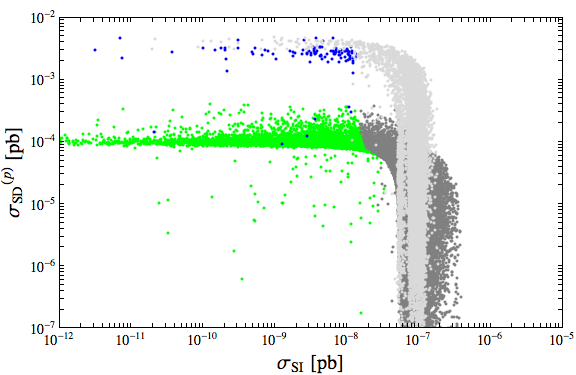}
\end{center}
\caption{\label{fig:SIvsSDMH140}Scatter plots of $\sigma_{\text{SD}}^{(p)}$ against $\sigma_{\text{SI}}$ depicting points with the correct relic density.  Shown are $m_{\nu_1} \leq 70 \text{ GeV}$ [top] and $m_{\nu_1} \geq 85 \text{ GeV}$ [bottom].  At bottom, the blue/light gray points represent $85 \text{ GeV} \leq m_{\nu_1} \leq 160 \text{ GeV}$ and green/darker gray represent $175 \text{ GeV} \leq m_{\nu_1} \leq 500 \text{ GeV}$; these mass ranges are chosen to avoid regions where $WW^\ast$ and $tt^\ast$ final states are expected to become important (see text for discussion).  In both plots, gray indicates points excluded by current direct detection limits.}
\end{figure}

In addition, the lack of signal events in XENON100 also implies new direct detection limits on $\sigma_{\text{SD}}^{(p,n)}$.  Based on the fact that the $\sigma_{\text{SI}}$ limits have improved by approximately a factor of 10 between XENON10 \cite{Aprile:2008rc} and XENON100,
we use the XENON10 spin-dependent limits 
to project that the $\sigma_{\text{SD}}$ limits will be $\mathcal{O}\left(10^{-3}\mbox{ pb}\right)$ and  $\mathcal{O}\left(10^{-2}\mbox{ pb}\right)$ for scattering off of neutrons and protons respectively.  The ratio $\sigma_{\text{SD}}^{(p)}/\sigma_{\text{SD}}^{(n)} \simeq 1.3$ for all points (resulting solely from the different couplings of the $Z$ to protons and neutrons).  Thus, we expect XENON100 limits on $\sigma_{\text{SD}}^{(n)}$ to be competitive with those from SIMPLE and Super-K on $\sigma_{\text{SD}}^{(p)}$ for $m_{\nu_1} \lesssim 200 \text{ GeV}$ (for higher masses, the significantly stronger limits from IceCube become relevant).  While at first glance it may appear that much of the parameter space is out of the reach of both present or near future direct detection, it is important to consider the 
correlation between $\sigma_\mr{SI}$ and $\sigma_\mr{SD}$.
This is represented in Fig.~\ref{fig:SIvsSDMH140}, which depicts the allowed points in the $\sigma_{\text{SD}}^{(p)}$ vs.~$\sigma_{\text{SI}}$ plane.

We see that in a large portion of the parameter space permitted by constraints on $\Omega h^2$, points have either a significant spin-independent or spin-dependent cross section.  
For heavier dark matter (with $m_{\nu_1} \ge 85 \text{ GeV}$), 
the majority of points lie in either a horizontal band at the top of the plot or a vertical band to the right.  The horizontal band consists of points for which the relic density is predominantly set by annihilation via $s$-channel $Z$ exchange, and these points correspondingly have the largest spin-dependent cross sections.  The vertical band contains points for which the dark matter annihilates predominantly via $s$-channel Higgs boson exchange, resulting in larger spin-independent cross sections.  The horizontal band is at lower values of $\sigma_{\text{SD}}^{(p)}$ for $175 \text{ GeV} \leq m_{\nu_1} \leq 500 \text{ GeV}$ than for $85 \text{ GeV} \leq m_{\nu_1} \leq 160 \text{ GeV}$ due to the opening of the $\nu_1 \nu_1 \rightarrow t t$ channel.  The $\nu_1 \nu_1 \rightarrow Z \rightarrow tt$ channel is significant, so its opening permits a smaller $\nu_1 \nu_1 Z$ coupling, yielding smaller spin-dependent cross sections.  The location of the vertical band is largely unchanged as the top threshold is crossed because the $\nu_1 \nu_1 \rightarrow h \rightarrow VV$ (where $V$ is $W$ or $Z$) channel dominantes the $\nu_1 \nu_1 \rightarrow h \rightarrow tt$ channel for $m_{\nu_1} \ge m_t$.  Notably, both spin-independent and spin-dependent searches are vital for probing this parameter space, as while many points have small $\sigma_\mr{SI}$ or $\sigma_\mr{SD}$, relatively few exhibit suppression of both.

Points that do have relatively small $\sigma_\mr{SI}$ \emph{and} $\sigma_\mr{SD}$ (those that do not clearly fall into a band) are those for which co-annihilation and $t$-channel annihilation to gauge bosons are particularly significant in the early universe.  This permits smaller couplings of the dark matter to the Higgs and $Z$ bosons, producing smaller spin-independent and -dependent cross sections.  In general points outside of, but near to, the bands are those for which $t$-channel processes are significant.  The masses of other dark sector particles are close enough to $m_{\nu_1}$ that $t$-channel exchange is not heavily suppressed, but sufficiently separated that co-annihilation is not  relevant in the early universe.  
As the masses of the dark sector particles become increasingly degenerate, $t$-channel annihilation processes increase in significance, and eventually co-annihilation becomes relevant.  The points further from both bands are those for which $t$-channel annihilation and co-annihilation are the dominant processes in setting the relic density, so $\sigma_{\text{SI}}$ and $\sigma_{\text{SD}}$ can be small (and in general must be to avoid over-annihilation).

For $40 \text{ GeV} \leq m_{\nu_1} \leq 70 \text{ GeV}$ (the upper plot in Fig.~\ref {fig:SIvsSDMH140}), there is no clear banding structure.  In this mass regime, lower spin-independent and spin-dependent cross sections can be achieved due to the presence
of the Higgs and $Z$ boson poles.
This allows the relic density to still be set by $s$-channel Higgs or $Z$ boson exchange but with significantly smaller $\nu_1 \nu_1 Z$ or $\nu_1 \nu_1 h$ couplings to compensate for the enhancement in the annihilation cross section due to the small propagator.  The contribution to the cross section from the propagator in the early universe goes as $(s - m_{h/Z}^2)^{-2} \simeq (4 m_{\nu_1}^2 - m_{h/Z}^2)^{-2}$, whereas for direct detection the propagator contribution goes as $m_{h/Z}^{-4}$.  As a result, enhancement of the annihilation cross section near a pole does not imply a similar enhancement of direct detection cross sections.  Points exhibiting this enhancement are numerous;
the dark matter need not be exactly on resonance to take advantage of a reduced  $s$-channeled propagator. Furthermore, the energies of the dark matter particles 
follow a Boltzmann distribution,
so for $m_{\nu_1} \lesssim m_Z/2, m_h/2$ some particles will have enough energy to utilize the resonance.

Thus, for fermionic WIMPs of this type and a relatively light Higgs boson, much of the parameter space is already excluded.  The remaining options that avoid exclusion are:
\begin{enumerate}
\item The dark matter mass allows annihilation through a Higgs or $Z$ boson that is enhanced due to the presence of an $s$-channel pole in the early universe.  This allows smaller couplings to the Higgs and $Z$ bosons, and suppressed spin-independent and spin-dependent cross sections respectively.
\item The dark sector masses are sufficiently close that dark matter annihilation in the early universe is predominantly due to $t$-channel processes or co-annihilation.  For many such models, direct detection is unobservable.
\item The dark matter coupling to the Higgs boson is small, suppressing $\sigma_{\text{SI}}$.  
The relic density is set by
$Z$ exchange, which generically leads to large spin-dependent cross sections.  Many of these models may be ruled out within the coming years  
by
direct detection experiments.  In particular, models 
with suppressed $\sigma_{\text{SI}}$ and $85 \text{ GeV} \leq m_{\nu_1} \leq 160 \text{ GeV}$ in which relic density is set by $s$-channel $Z$ exchange are already beginning to be excluded by spin-dependent direct detection experiments.
\end{enumerate}
In each of these scenarios, some tuning of the parameters is required.  In the first case, it is necessary to have $m_{\nu_1} \lesssim m_Z/2$ or $m_h/2$.  For case 2, the masses of the dark sector particles must be nearly degenerate, $\Delta m \alt T_{fo} \simeq m/20$, and $\sigma_{\text{SI}}$ and $\sigma_{\text{SD}}^{(p)}$ must also be fairly small.  
This usually
requires $M_S \simeq M_D$, and small $\lambda$ and $\lambda'$.
In the final case, for a given value of $\lambda$, $\lambda'$ must be tuned to be approximately $\lambda'_{\text{crit}}$.  At present, the required a tuning is mild, at the level of approximately ten percent; setting $\lambda'$ to within $\sim 10\%$ of $\lambda'_{\text{crit}}$ will suppress $\sigma_{\text{SI}}$ by a factor of $\mathcal{O}(10^2)$.   

The allowed parameter space will become even more restricted with imminent developments in dark matter detection experiments.  A one-ton Xe experiment could potentially improve bounds on spin-independent cross section by orders of magnitude.  For points with suppressed $\sigma_{\text{SI}}$, improvements in experiments that probe $\sigma_{\text{SD}}$ will be very important.  Projected limits from the COUPP experiment \cite{Behnke:2008zza} are on the order of $\sigma_{\text{SD}} \sim 10^{-3} - 10^{-4} \text{ pb}$ for dark matter masses between 10 and 500 GeV.  In addition, experiments other than those that focus on direct detection of dark matter may begin to play a role.  For instance, recent work has shown that bounds on monojet events the LHC on $\sigma_\mr{SD}$ are rapidly becoming comparable to direct detection bounds \cite{Rajaraman:2011wf}; however, these currently only apply if the operator mediating direct detection is effectively parameterized by a contact operator at the LHC.  Here, where Z boson exchange is relevant, a preliminary investigation indicates that the collider bounds are significantly degraded.  A more promising probe is the DeepCore extension to IceCube, which should also provide stringent limits on $\sigma_{\text{SD}}$ for dark matter in this mass range \cite{Wiebusch:2009jf}.  A recent study \cite{Barger:2011em} has found that the expected atmospheric background rate for muon events DeepCore is approximately 2.3 events per year.  This informs the estimate that the dark matter annihilations in the Sun must yield approximately 10 muon events per year for discovery.  
We can thus approximate the capture and annihilation rates in the sun necessary to produce this required number of events, and consequently the spin-dependent cross sections that we expect to be probed by DeepCore. We rescale points A and D from \cite{Barger:2011em}, accounting for the dominant mass dependent effects.  Doing so, we find that for a dark matter candidate annihilating primarily to $\tau \tau$ and $b b$ (for $m_W \lesssim m_{\text{DM}} \lesssim m_t$) or $tt$ (for $m_{\text{DM}} \gtrsim m_t$), the approximate $\sigma_{\text{SD}}^{(p)}$ required for discovery rises from $\sim 2 \times 10^{-5} \text{ pb}$ for a 100 GeV dark matter candidate to around $10^{-4} \text{ pb}$ for a 500 GeV dark matter candidate.  This is comparable to, although slightly less optimistic than, the projected limits given in \cite{DeepCoreTalk}, which assume a lower energy threshold will be attainable. For points with these relatively high spin-dependent cross sections, annihilation rates are sufficiently high that the WIMPs in the sun are in equilibrium.

If no hint of dark matter is seen at DeepCore, we expect the experiment will severely limit the available parameter space for the fermionic singlet-doublet model in the case of $m_{\nu_1} \ge m_W$.  For $m_W \leq m_{\nu_1} < m_t$, points with suppressed $\sigma_{\text{SI}}$, and relic density and neutrino spectrum set by annihilation via an $s$-channel $Z$ (to $W W$ in the early universe and to $\tau \tau$, $bb$ in the Sun and Earth - those in the horizontal blue band of Fig. \ref{fig:SIvsSDMH140}) could soon be readily excluded by a combination of  direct detection experiments sensitive to spin-dependent couplings and DeepCore.  In the case of $m_{\nu_1} \gtrsim m_t$, points with suppressed $\sigma_{\text{SI}}$, with correct relic density and neutrino spectrum set by annihilation to $t t$ via $s$-channel $Z$ exchange (the horizontal green band of Fig. \ref{fig:SIvsSDMH140}) generally exhibit spin-dependent cross sections that are comparable to (if not slightly greater than) the expected DeepCore limits after one year of running.  Consequently, for $m_{\nu_1} \ge m_W$, it may soon be the case that scenario 2 is the only viable option for avoiding experimental constraints.  For $m_{\nu_1} < m_W$, the situation is less clear: there are a number of points with lower $\sigma_{\text{SD}}$, and the annihilation of lighter dark matter will yield a softer neutrino spectrum, so the prospects for detection will depend significantly on the precise muon detection energy threshold achieved by DeepCore.  Direct detection experiments will still be important in this range.
 
One clear take-away from this analysis is that a combination of spin-independent and spin-dependent experiments will be necessary to effectively probe the variety of dark matter models; neither one will be sufficient on its own to eliminate the majority of the parameter space for this model of dark matter.  Furthermore, given the correspondence between direct detection and annihilation in the early universe, measurements from both types of experiment may be vital to determine the properties of a dark matter particle.

\subsection{Heavier Higgs bosons}\label{sec:HeavierHiggsBosons}

We now consider how the situation changes when we increase the mass of the Higgs boson.  
Within the Standard Model, recent ATLAS \cite{leptonphotonATLAS2011} and CMS \cite{leptonphotonCMS2011} results disfavor 
most of
the range $150 \lesssim m_h \lesssim 450$ GeV.   For moderate values of the Higgs boson mass, however, 
LHC production
cross sections not much below the Standard Model 
rate
are allowed.  In the model with mixed singlet-doublet fermion dark matter, there is 
the possibility that the Higgs boson decays invisibly 
 into pairs of neutral ${\mathbb Z}_2$-odd fermions with an appreciable branching ratio
allowing evasion of the ATLAS and CMS 95\% CL limits.   
However, for a Higgs boson in this mass range, invisible decays compete with decays to $WW$, so achieving even an $\simeq 10\%$ branching ratio requires large couplings to the dark sector.  This  leads to spin-independent direct detection cross sections that are already in excess of XENON100 bounds.  If we repeat the scan of Fig.~\ref{fig:SIDmassMH140} for a 200 GeV Higgs boson (including the appropriate 
constraint
on $\Delta T$) with the additional requirement that the Higgs boson has a $\geq 10 \%$ branching ratio to dark sector particles, we 
find no allowed points.
This is true for Higgs bosons in the entire ATLAS/CMS exclusion range as well.

A Higgs boson heavy enough to evade LHC searches, $m_h\gsim 450$ GeV, requires a large positive contribution to the $T$ parameter from new physics 
in order
to be consistent with precision electroweak data.  As has been pointed out in \cite{D'Eramo:2007ga,Enberg:2007rp}, it is possible for this correction to arise from the effects of the dark sector itself. To explore the viable parameter space for a heavy Higgs, we repeat the scans that produced Figs.~\ref{fig:SIDmassMH140} and \ref{fig:SIvsSDMH140}, this time with $m_h = 500$ GeV, and with the dark sector's contribution to the $T$ parameter constrained to be in  the range
\begin{equation}
0.16 < \Delta T < 0.40.
\end{equation}
We scan over the same parameter ranges as for the $m_h$=140 GeV case.  While we assume the new $\Delta T$ contribution arises from the dark sector itself, it is possible to imagine a more baroque model where the additional  new physics contributes to $\Delta T$.   In this case, an increase in  $m_h$ can generically be used to suppress $\sigma_{\text{SI}}$.   We do not focus on this case here as it is phenomenologically straightforward.
\begin{figure}
\begin{center}
 \includegraphics[width=0.75\textwidth]{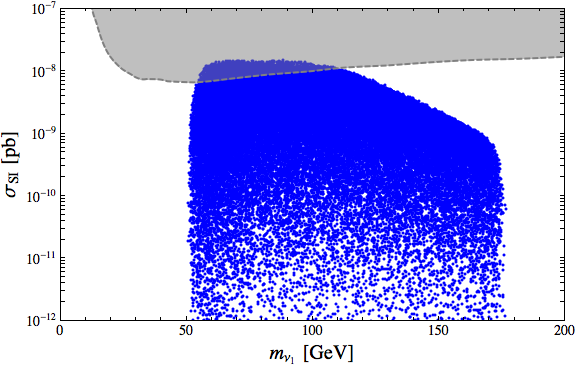}
  \includegraphics[width=0.75\textwidth]{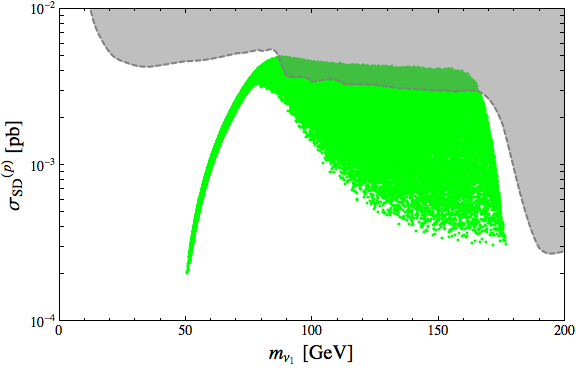}
\end{center}
\caption{Plots of spin-independent [top] and spin-dependent [bottom] cross section against dark matter mass for $m_h = 500$ GeV.  Exclusion contours are as in Fig.~\ref{fig:SIDmassMH140}.}
\label{fig:fermion_heavyhiggs_DDvsM}
\end{figure}

In Fig.~\ref{fig:fermion_heavyhiggs_DDvsM} we show the results for spin-independent and spin-dependent cross sections versus dark matter mass.   
At tree-level the spin-independent cross section depends on the $\nu_1 \nu_1 h$ coupling,  which can be arbitrarily small given the potential cancellations discussed in 
Sec.~\ref{sec:suppressed}.  
For  $m_{\nu_1}< m_W$,  
$Z$ exchange  
regulates the relic abundance.
For dark matter masses above  $m_W$,  Higgs boson mediated annihilations to $WW$  can instead set the abundance, but the possibility of using the $Z$ coupling alone to do so persists in this regime as well. Regardless of whether $m_{\nu_1}$ lies below or above $m_W$, it is therefore possible to tune the $\nu_1 \nu_1 h$ coupling away and still achieve a realistic relic abundance.  Although the great majority of points have spin-independent cross sections within roughly two orders of magnitude of current limits, points with tiny spin-independent cross sections consequently show up in the full mass range 
from $\sim 50-170$ GeV.

An important feature 
of both plots in
Fig.~\ref{fig:fermion_heavyhiggs_DDvsM} is that no points show up for  $m_{\nu_1} > m_t$.
 Our requirement that  the dark sector produces a large $\Delta T$ (which goes parametrically as $(\lambda^2 - \lambda^{\prime 2})^2$) forces $\lambda$ and $\lambda'$ to have very different magnitudes, which in turn means that the $\nu_1 \nu_1 Z$ coupling
will generally be significant.  Since $Z$ boson mediated annihilations to $t {\bar t}$ do not suffer from $p$-wave suppression,
achieving the correct relic density when the $\nu_1 \nu_1 \rightarrow Z \rightarrow t t$ channel is open requires a small $\nu_1 \nu_1 Z$ coupling.  
Hence, it is impossible to simultaneously satisfy the requirement of large $\Delta T$ and 
 the constraint on the relic density, thereby
prohibiting points with $m_{\nu_1} > m_t$.  If we were to relax our requirement that $\Delta T$ come from the dark sector, 
smaller values of the  $\nu_1 \nu_1 Z$  coupling would be possible and the $m_{\nu_1}> m_t$ region would open.
\begin{figure}
\begin{center}
 \includegraphics[width=0.75\textwidth]{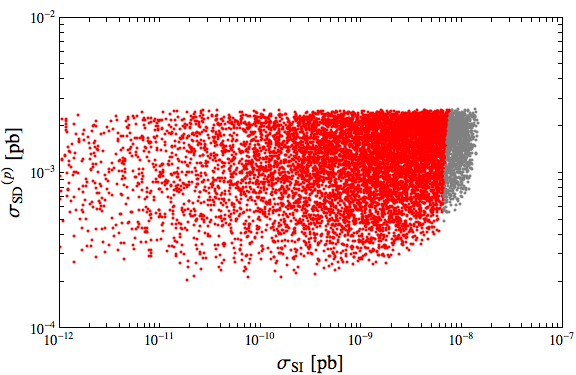}
  \includegraphics[width=0.75\textwidth]{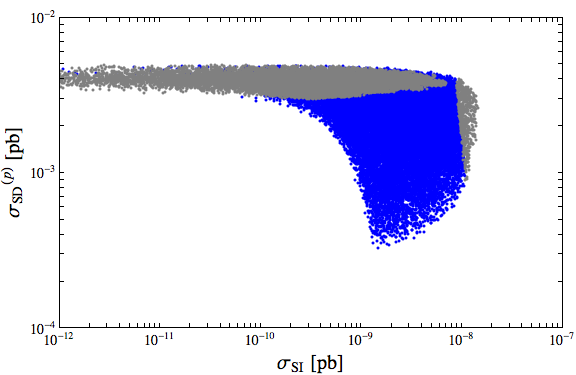}
\end{center}
\caption{Scatter plots of  $\sigma_{\text{SD}}^{(p)}$ against $\sigma_{\text{SI}}$ depicting points with the correct relic density, for $m_h = 500$ GeV.  Shown are $m_{\nu_1} \leq 70 \text{ GeV}$ [top] and $m_{\nu_1} \geq 85 \text{ GeV}$ [bottom].   In both plots, gray represents points already excluded by direct detection experiments.}
\label{fig:fermion_heavyhiggs_SIvsSD}
\end{figure}

Next we turn
our attention to the second plot in Fig.~\ref{fig:fermion_heavyhiggs_DDvsM}.  
For $m_{\nu_1}< m_W$, where annihilation through an $s$-channel $Z$ sets the abundance,  the $ \nu_1 \nu_1 Z$ coupling  required to obtain the correct relic 
density
gets smaller as $m_{\nu_1}$ approaches $m_Z/2$ from above, due to the enhancement 
from
the $s$-channel propagator.  This results in smaller spin-dependent  cross sections.  When $m_{\nu_1}$ gets sufficiently close to $m_Z/2$, the propagator enhancement becomes so large that it becomes impossible to find $\lambda$ and $\lambda'$ values such that $\Delta T$ is large enough enough while $ \nu_1 \nu_1 Z$ is simultaneously small enough 
to acheive a realistic relic abundance.
This explains why no points are realized for $m_{\nu_1} \lesssim 50$ GeV for both plots in Fig.~\ref{fig:fermion_heavyhiggs_DDvsM}.  Analogously to the $m_t$ cutoff discussed in the previous paragraph, the cutoff at around 50 GeV is tied to our $\Delta T$ requirement.

At larger values of $m_{\nu_1}$ the spin-dependent cross section is rather large, $\sigma_{\text{SD}}^{(p)} \gsim 4 \times 10^{-3}$ pb,  for points where the abundance is set by the coupling to the $Z$.   
Note that for these larger masses,
the $WW$ and $ZZ$ final states are also available.
Therefore,
a non-trivial contribution to annihilation from Higgs boson exchange is possible, and a realistic abundance may be found for smaller $Z$ couplings. 
This yields points
with smaller spin-dependent cross sections, although these cross sections are non-vanishing because the $\Delta T$ requirement prevents the $\nu_1 \nu_1 Z$ coupling from being extremely suppressed.  

These observations are also relevant for
understanding the plots of spin-dependent versus spin-independent cross sections shown in Fig.~\ref{fig:fermion_heavyhiggs_SIvsSD}.  
As mentioned above, $Z$ exchange necessarily regulates the abundance for masses below $m_W$  --- this places a minimum value on the spin-dependent cross section of $\sigma_{\text{SD}}^{(p)} \sim 2 \times 10^{-4}$ pb.
Even if there is a delicately
canceled $\nu_1 \nu_1 h$ coupling, the spin-dependent cross section will be large enough to be seen at upcoming experiments.  The second plot in Fig.~\ref{fig:fermion_heavyhiggs_SIvsSD} shows that this is also
true for larger $m_{\nu_1}$ values. 
For this higher mass region the effect is more pronounced, with spin-independent cross sections smaller than $10^{-10}$ pb requiring spin-dependent cross sections $\gtrsim 3 \times 10^{-3}$ pb.  Consequently, many of these points 
are excluded by current experimental bounds.   In this high mass region, if the Higgs boson coupling is suppressed, there is no pole enhancement for $Z$-mediated annihilation so we must regulate the abundance with a ``full-strength" $Z$ coupling, producing larger spin-dependent cross sections for points with suppressed $\sigma_\mr{SI}$ than in the low mass region.

As for the case of a light Higgs boson, DeepCore and direct detection experiments should be sufficiently sensitive to probe the points with $m_W \leq m_{\nu_1} < m_t$ and suppressed $\sigma_{\text{SI}}$.  Furthermore, since in this case there is a floor on $\sigma_{\text{SD}}$ for $m_{\nu_1} < m_W$, these experiments could also have interesting implications for lighter dark matter.  Consequently, in this regime the most difficult points to probe may be those for which $m_{\nu_1} \simeq m_t$. 
As the $\nu_1 \nu_1 \rightarrow Z \rightarrow t t^\ast$ annihilation channel begins to turn on,
a smaller $\nu_1 \nu_1 Z$ coupling
can be allowed
 (and thus a smaller $\sigma_{\text{SD}}$), 
implying that these
points are more difficult to probe.

In summary, we see that for $m_h = 500$ GeV,  the vast majority of points will be probed  through their spin-independent cross sections once the experiments improve their reach by about  two orders of magnitude.  Even points with unusually small spin-independent cross sections should be probed through their spin-dependent cross sections in the near future.  These conclusions are sensitive to our assumption that the dark sector produces a large $\Delta T$.

\section{The scalar model}\label{sec:scalar}
We now consider the analogous model where the fermions are replaced with scalars. 
A simple candidate model of dark matter, 
it displays a broader range of phenomenology than the simplest model 
 of scalar WIMP dark matter where the abundance of a real singlet scalar 
 is set via a quartic coupling to the Higgs field \cite{Silveira:1985rk,McDonald:1993ex,Burgess:2000yq}.  
While scalar singlet dark matter is not yet ruled out, 
 future direct detection experiments 
may soon begin to eliminate this simplest model for lighter Higgs boson masses.  
Consequently, it 
is worthwhile to consider whether extending such a model to include an additional doublet can potentially allow for evasion of future direct detection bounds.

We introduce a real scalar singlet $S$ and a complex doublet $\Phi$ (with hypercharge $1/2$) 
and the Lagrangian 
\begin{eqnarray}
\Delta \mathcal{L} & = & 
 D_\mu \Phi^\dagger D^\mu \Phi - m_D^2 \Phi^\dagger \Phi + \frac{1}{2} (\partial_\mu S)^2 - \frac{m_S^2}{2} S^2 - g (S \Phi^\dagger H + \text{h.c.}) \nonumber \\
&& - \frac{\lambda_S}{2} S^2 H^\dagger H - \lambda_1 (H^\dagger H) (\Phi^\dagger \Phi) - \lambda_2 \left((\Phi^\dagger H)^2 + \text{h.c.}\right) - \lambda_3 (\Phi^\dagger H) (H^\dagger \Phi),
\label{eqn:SDScalarsLagrangian}
\end{eqnarray}
where $SU(2)$ indices are contracted within parentheses, and the doublet is
\begin{equation}
\Phi \equiv \left(
\begin{array}{c}
\phi^+ \\ 
\frac{1}{\sqrt{2}} \left(\phi^0 + i A^0\right)
\end{array}
\right).
\label{eqn:ScalarDoubletDef}
\end{equation}
We neglect other possible allowed couplings containing only dark sector particles that are not relevant to the
dark matter phenomenology, 
e.g.~$S^2 (\Phi^\dagger \Phi)$. For non-zero trilinear coupling $g$,  the singlet and the doublet mix when the Higgs boson takes on its vev.  The resulting dark matter is:
\begin{equation}
X_1= \cos{\theta} \,  S + \sin{\theta}  \, \phi^{0}.
\end{equation}
We denote the orthogonal neutral scalar as $X_{2}$.
  
In contrast to the fermion case, annihilations through the Higgs boson can be present without inducing mixing, for instance due to the presence of the $S^2 (H^\dagger H)$ coupling.
In the presence of non-zero mixing, the coupling to the Higgs boson is given by:
\begin{eqnarray}
\mathcal{L} & \supset & - (\lambda_S v \cos^2 \theta + \lambda_{123} v \sin^2 \theta - 2 g \sin\theta \cos\theta) X_1^2 h \equiv - A_{\text{eff}} X_1^2 h,
\label{eqn:Aeff}
\end{eqnarray}
where we have introduced the effective coupling of the neutral doublet scalar to the Higgs boson $\lambda_{123} \equiv \lambda_{1} + 2 \lambda_{2} + \lambda_{3}$.  

\begin{figure}
\vspace{-4in}
\hspace{-1in}
\includegraphics[width=\textwidth]{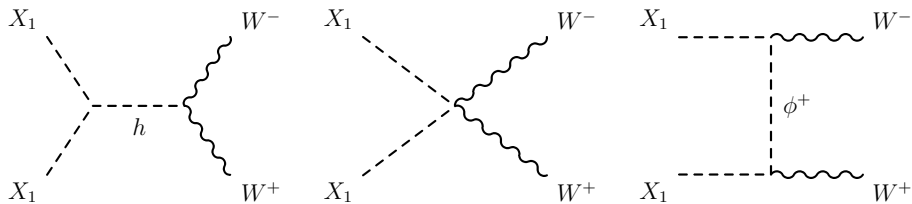}
\vspace{-3.8in}
\caption{\label{fig:scalarDiagrams}The dominant annihilation processes for singlet--doublet scalar 
dark matter
 in the regime $m_{X_1} > m_{W}$.}
\end{figure}

The dominant processes that contribute to early universe annihilation in this model (for $m_{X_{1}} > m_{W}$) are shown in Fig.~\ref{fig:scalarDiagrams}.  For masses beneath the $W$-boson mass, the relic abundance is essentially determined by the $s$-channel Higgs boson exchange diagram, with coupling $A_\mr{eff}$ and a $b \bar{b}$  final state.

\begin{figure}[h]
\includegraphics[width=.75\textwidth]{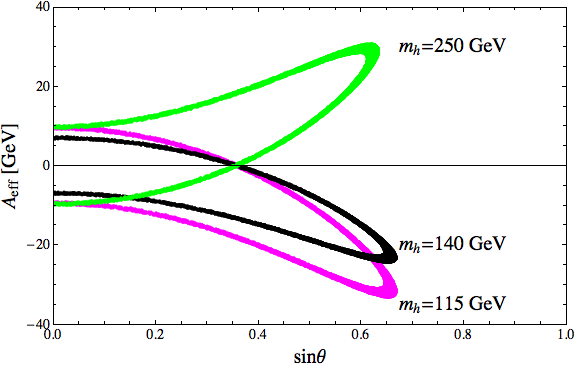}
\includegraphics[width=.75\textwidth]{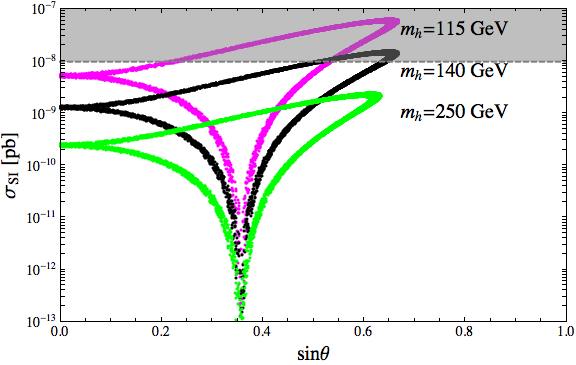}
\caption{\label{fig:ScalarVaryHiggs} In the top panel, we show the coupling to the Higgs boson, $A_\mr{eff}$ (see 
Eq.~(\ref{eqn:Aeff})), 
needed to achieve the correct relic density as a function of the mixing angle $\sin \theta$.  Regions are shown for three different Higgs boson masses: $m_{h} = 115$ GeV,  $m_{h} = 140$ GeV, and $m_{h} = 250$ GeV.  The dark matter mass is fixed, $m_X = 95$ GeV, and all $\lambda_i=0$, i=1,2,3.  In the bottom panel,  we plot 
 $\sigma_\mr{SI}$  vs. $\sin{\theta}$, from top to bottom, $m_h= 115, 140, 250$ GeV.  The shaded region corresponds to the 
XENON100
exclusion for this mass \cite{Aprile:2011hi}. 
}
\end{figure}

It is instructive to examine the region of correct thermal relic density in the $\sin \theta - A_\mr{eff}$ plane (the upper panel of Fig.~\ref{fig:ScalarVaryHiggs}).  In this figure, we have shown the allowed region for three choices of Higgs boson mass, $m_{h} =115, 140, 250$ GeV
with $m_{X_{1}}=95\mbox{ GeV}$
and $m_D = 125$ GeV.  The $\lambda_{i} =0$, for $i=1,2,3$.  
Setting these couplings to zero ensure the absence of any co-annihilation, a possibility we will revisit below.
Scalars contribute less to the $T$ parameter than fermions
with similar strength couplings, 
so we do not require internal $\Delta T$ to compensate for heavier Higgs boson masses.   Moreover, for $\lambda_3 = 2 \lambda_2$ a custodial $SU(2)$ is maintained in the new sector, such that $\Delta T$ vanishes for all of the points shown in these plots (as $\lambda_3 = 2 \lambda_2 = 0$).\footnote{This custodial symmetry can be made manifest as follows. Write
$\Omega_H =(\tilde{H} H)$ 
which
transforms under $SU(2)_L \times SU(2)_R$ as $\Omega_H \rightarrow L \Omega_H R^\dagger$.  The $\Phi$ doublet has the same quantum numbers as the Higgs doublet, so we can have an analogous $\Omega_\Phi$ that transforms identically.  Then, for $\lambda_3 = 2\lambda_2$, we can write $\Delta \mathcal{L} \supset - g S\; \text{tr}(\Omega_\Phi^\dagger \Omega_H) - \frac{\lambda_1}{4} \text{tr}(\Omega_\Phi^\dagger \Omega_\Phi) \text{tr}(\Omega_H^\dagger \Omega_H) - \lambda_2 [\text{tr}(\Omega_\Phi^\dagger \Omega_H)]^2$, and the custodial symmetry is explicit.}

At $\sin \theta=0$, 
for $m_S < m_D$, 
we recover the ``pure singlet model" (remove all terms with  $\Phi$ from Eq.~(\ref{eqn:SDScalarsLagrangian})) and its attendant value of $|A_\mr{eff}|$.
Moving away from $\sin \theta=0$, other processes begin to contribute to 
$X_1 X_1 \rightarrow W^+ W^-$.  
The dominant effect is due to
the direct 4-point vertex (the middle diagram in Fig.~\ref{fig:scalarDiagrams}); the $t$-channel exchange is usually smaller.  The presence of these additional diagrams  requires a new value of $A_\mr{eff}$ to maintain the correct relic abundance.  Notably, there exists a value of $\sin \theta$ for which the correct relic density is maintained only via the gauge interactions, and the contribution from the Higgs boson vanishes ($A_\mr{eff}$=0). At this point, the spin-independent detection cross section
plummets.  This explains the deep trough in the lower panel of Fig.~\ref{fig:ScalarVaryHiggs}.   Once again, it should be noted that where tiny cross sections appear here (and elsewhere in this section), loop induced effects which we have neglected in our numerical studies would be relevant.  

We now discuss the interplay between
the contributions from
Higgs boson exchange 
and the 4-point diagram to $X_1 X_1 \rightarrow W^+ W^-$
in more detail.   The interference between these two diagrams can be constructive or destructive.  This depends on two factors:  the sign of $A_\mr{eff}$ and the size of the Higgs boson mass.  The latter  (in combination with the dark matter mass) sets the sign of the $s$-channel propagator. Examining the lower panel of Fig.~\ref{fig:ScalarVaryHiggs}, there is a plateau of relatively large $\sigma_\mr{SI}$ values.  There
the relic density is set dominantly via $s$-channel Higgs boson exchange.   The 4-point diagram makes a subdominant contribution that interferes \textit{destructively} with the Higgs diagram.  Consequently, 
the $|A_\mr{eff}|$ must be increased to maintain the correct relic abundance.  
In the top panel, this can be seen for the lower (upper) branches of the 
curve
for $m_h=115,140$ (250) GeV.  
Due to the increased size of $|A_\mr{eff}|$,
direct detection cross sections  are greater than those found in the model with no doublet at all.  For sufficiently large values of $\sin \theta$, there is another possibility exhibiting the reverse situation:  annihilation may be dominated by the four--point diagram, with a subdominant $s$-channel Higgs boson contribution that  interferes destructively.  In the top panel, this corresponds to the segment that extends from $|A_\mr{eff}| =0$ up to the tip of the  
curve.  In the lower panel, this segment extends from the trough up to values of peak cross section at large $\sin \theta$.  The tip of the 
curve is characterized by points at which 
the destructive interference between the four-point and the $s$-channel Higgs boson diagrams is most severe.  In this region, other processes such as $t$-channel charged scalar exchange, annihilation via an $s$-channel Higgs boson to heavy quarks (for instance, $t t$ for $m_{\nu_1} > m_t$) or annihilation to Higgs boson pairs (for $m_{\nu_1} > m_h$) can play significant roles.  Finally, there is a region where the interference is \textit{constructive}.  In the upper plot, this segment runs from $\sin \theta=0$ (where only Higgs boson exchange contributes)  out to ($\sin \theta, A_\mr{eff}$) = (0.35, 0), where only the four-point diagram contributes.  In the lower plot, this 
explains 
the lower left portion of the triangular region. 

To summarize, 
the presence of 
additional contributions to the
 $X_1 X_1 \rightarrow W^+ W^-$ annihilation channel can either increase  or decrease  the direct detection cross section with respect to a 
dark matter
candidate that relies on annihilation via a Higgs boson alone.  XENON100 has already begun to probe this model for lower values of the Higgs boson mass.
To explore
the 
achievable
direct detection cross sections in this model, we performed a scan over all parameters with the ranges:
$10 \text{ GeV} \leq m_{X_1} \leq 500 \text{ GeV}$, $80 \text{ GeV} \leq m_D \leq 1 \text{ TeV}$, $\left|\lambda_i\right| \leq 1$, $0
\leq g \leq v$.  We imposed the same $\Delta T$ requirements as in fermion case with a light Higgs boson,\footnote{As alluded to previously, this prohibits very few points due to the difficulty of achieving large $\Delta T$ contributions from scalars.  However, we include this requirement for consistency.} and required that the sum of each scalar mass and the pseudoscalar mass be greater than $m_Z$ (to avoid $Z$-width constraints).  Note, there is a possibility that the dark matter might be quite light, $\alt$ few GeV, consistent with current direct detection bounds.  In this case, the phenomenology is essentially that of the pure singlet, coupled to a Higgs boson.  This window was studied recently in \cite{He:2011de}, see also \cite{Raidal:2011xk}.

\begin{figure}[h]
\includegraphics[width=\textwidth]{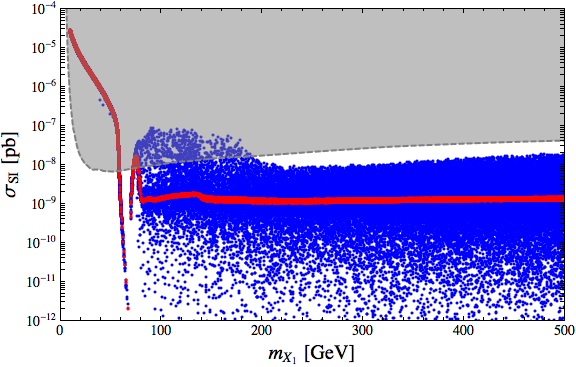}
\caption{\label{fig:ScalarOverview} $\sigma_\mr{SI}$ vs. $m_{X_1}$ for the scalar singlet doublet model.  The Higgs boson mass is $m_{h} = 140$ GeV.  Superimposed is the narrower band that corresponds to the pure singlet model.  Also shown is the exclusion region from XENON100  \cite{Aprile:2011hi}.}
\end{figure}

The result is shown in Fig.~\ref{fig:ScalarOverview}.  Superimposed on this plot is a scan over the pure singlet model.
In the singlet model, all dynamics are controlled by the Higgs--dark matter coupling.  
The precise measurement of the dark matter relic abundance determines  $\lambda_S$, which in turn determines $\sigma_\mr{SI}$, resulting in the thin band in the figure.  The addition of the doublet allows deviations from this curve.  Points approximately along the curve are those whose relic abundance is set by the coupling to the Higgs boson, $A_{\text{eff}}$, of Eq.~(\ref{eqn:Aeff}).  For $m_{X_1} > m_{W}$ other
channels can now contribute to annihilation in the early universe, and the firm connection between (Higgs boson mediated) direct detection and cosmology is broken.  Nevertheless, many of the points in the plot will be probed by a future generation of direct detection experiments.  Features in the plot can also be observed at the $t t$ and $hh$ thresholds, where new final states open up.  Some of the points with the lowest $\sigma_\mr{SI}$ are due to the 
minimum exhibited in Fig.~\ref{fig:ScalarVaryHiggs} (where four point diagram $X_1 X_1 \rightarrow W^+ W^-$ sets the relic density).

$\sigma_\mr{SI}$ can also be suppressed if co-annihilation is relevant. Since
the dark matter is a real scalar, it does not possess diagonal couplings with the $Z$ boson.  
Any mass splitting between $A^{0}$ and $X_1$ which is $\agt 100$ keV avoids an enormous (lethal) $Z$-boson mediated spin-independent cross section. If the splitting is close to this value, the scattering is inelastic
\cite{TuckerSmith:2001hy}.  Since we are considering $g \gsim \mathcal{O}$(GeV), it is unlikely that such a small splitting will be realized.  However, it is possible that the pseudo-scalar may have mass sufficiently close to the scalar so that this off-diagonal coupling is relevant for
setting the relic density in the early universe via co-annihilation.  Similarly, the charged scalar, $\phi^{+}$ may co-annihilate with the 
dark matter
via the $W$ boson.   

\begin{figure}[h]
\includegraphics[width=.74\textwidth]{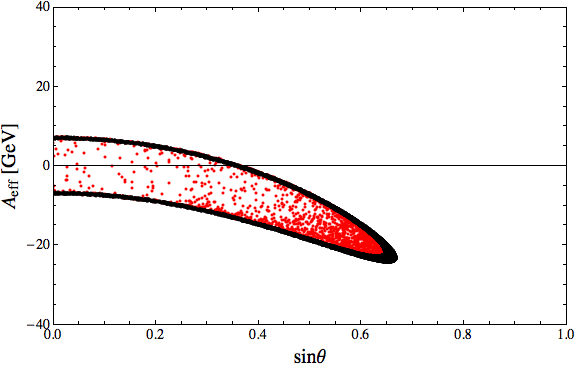}
\includegraphics[width=.74\textwidth]{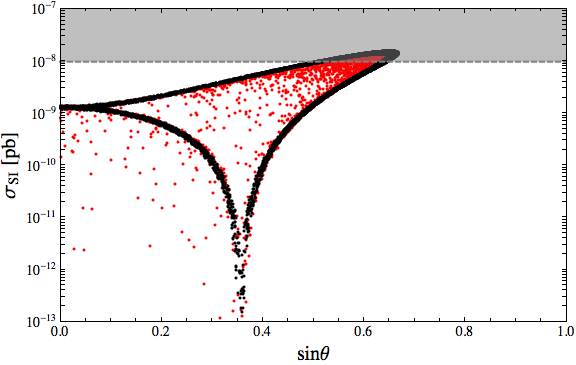}
\caption{\label{fig:ScalarCoannihilation}
The coupling to the Higgs boson, $A_\mr{eff}$ (see Eq.~(\ref{eqn:Aeff})), needed to achieve the corrrect relic density as a function of the mixing angle $\sin \theta$ (top).  The Higgs boson and dark matter masses
are fixed:  $m_{h} = 140$ GeV and $m_{X_1} = 95$ GeV.  Unlike Fig.~\ref{fig:ScalarVaryHiggs}  we allow  $\lambda_i \neq 0$, $ i=1,2,3$ but fix $m_{A^0} = 125$ GeV (see text for further discussion).  Points interior to the curve illustrate the possible relevance of co-annihilation.  At bottom,  we plot  $\sigma_\mr{SI}$
vs.~$\sin{\theta}$.   When compared to Fig.~\ref{fig:ScalarVaryHiggs}, there are points with  reduced $\sigma_\mr{SI}$.  The shaded region corresponds to the 
XENON100 exclusion for this mass \cite{Aprile:2011hi}.}
\end{figure}

To demonstrate the possible relevance of co-annihilation, we again examine the $A_\mr{eff} - \sin \theta$ plane, 
while relaxing the condition that the $\lambda_{i}=0$.
For concreteness, we choose a combination of $\lambda_{i}$ to allow the possibility that $m_{\phi^+} \approx m_{X_{1}}$, but we leave the pseudo-scalar mass fixed at $m_{A^0} = m_{D}= 125$ GeV.  The dark matter mass is again fixed at 95 GeV and $m_{h} = 140 \text{ GeV}$.  With respect to the analogous upper plot in  Fig.~\ref{fig:ScalarVaryHiggs},  we notice the possibility of points within the interior of the curve.
These are precisely the points where co-annihilation and $t$-channel exchange are relevant, and a smaller coupling to the Higgs boson may be accommodated.  For direct detection, the lower panel of Fig.~\ref{fig:ScalarCoannihilation}, there is the possibility of points with reduced detection cross-sections and small $\sin \theta$.

Finally, we note that a (nearly) pure doublet scalar can yield the correct
relic density.  All that is needed is a tiny splitting ($\agt 100$ keV) between the scalar and pseudo scalar state to avoid the enormous $Z$-boson mediated direct detection signals.  This can be accomplished via a tiny mixing with the singlet.  In this case, the right relic density is achieved for $m_{D} \simeq 500$ GeV.  Unfortunately, the direct detection cross section will be tiny in this case.  It might be possible to eventually observe an indirect detection signal.

\section{Conclusions}
We have explored models of strictly weakly interacting dark matter; specifically, dark matter whose annihilation, spin-independent and spin-dependent cross sections are controlled by the $W$, $Z$ and Higgs bosons.  Since
the neutral component of a pure electroweak doublet with full-strength coupling to the $Z$-boson has a fatally high direct detection cross section, we have considered the case in which these electroweak doublet couplings are diluted by mixing with a sterile state.
This singlet--doublet model serves as a proxy for strictly weakly interacting dark matter.  Other similar models are possible, such as mixing active dark matter in other representations of $SU(2)_L$ with a Standard Model singlet.  However, the singlet-doublet
model is particularly appealing since
it allows mixing between the active and sterile states to arise from renormalizable couplings to the Higgs field.
We have analyzed this type of model for 
the case where the new dark sector particles are fermions, and where they are scalars.  
These models subsume other models of
weakly-interacting mixed singlet-doublet dark matter, such as a mixed Bino-Higgsino state in supersymmetric extensions of the Standard Model.

We find that, for the case of both the fermion and scalar, current direct and indirect detection experiments are already beginning to probe the parameter space consistent with the required thermal relic density of $0.1053 \leq \Omega h^2 \leq 0.1193$.  Furthermore, near term
experiments should be capable of probing the majority of the parameter space, leaving only a few specific regions intact.  In much of the parameter space, the sizable couplings between the dark matter and the Higgs and/or $Z$ bosons required to achieve sufficient dark matter annihilation in the early universe imply correspondingly large spin-independent and/or spin-dependent cross sections, respectively.  

For a fermionic singlet--doublet WIMP, the prospects for discovery or exclusion are very optimistic.  While it is possible to suppress either $\sigma_\mr{SI}$ or $\sigma_\mr{SD}$ in the context of this model, the requirement of sufficient dark matter annihilation in the early universe makes suppressing both
cross sections
 extremely difficult.  Notably, this means that both $\sigma_\mr{SI}$- and $\sigma_\mr{SD}$-based dark matter detection experiments will be vital for discovering or excluding this class of models.  As spin-independent and spin-dependent limits improve, for instance with the advent of a one-ton XENON experiment and the DeepCore extension to IceCube, 
the most viable options for evading direct detection bounds 
are limited if the Higgs boson is light: 
either the annihilation in the early universe is enhanced by a small $s$-channel propagator (due to the Higgs or $Z$ boson poles) or coannihilation occurs.   

A heavy Higgs boson is also an option for avoiding
$\sigma_\mr{SI}$ limits.
However,
recent ATLAS and CMS limits have constrained ``heavy'' to 
imply
$m_h \gtrsim 450 \text{ GeV}$ for a Standard Model-like Higgs boson.  In this case, the large contribution to the $T$ parameter from the 
Higgs boson will require cancellation for consistency
with electroweak precision constraints. Such a contribution could come from the dark sector.
In the case of the fermionic singlet--doublet model this 
implies spin-dependent cross sections well within the reach of future experiments. 

The scalar model also exhibits sizable spin-independent cross sections in much of the parameter space.  
If the model is not discovered in the near future,  coannihilation or enhanced $s$-channel propagators again
provide options for avoiding direct detection limits.  For scalars, however, the is another option: 
$\sigma_\mr{SI}$ can be heavily suppressed while
the correct relic density is achieved by a sizable four-point $XXVV$ (with $V$ as $W$ or $Z$) coupling.  There is no appreciable $\sigma_\mr{SD}$ in this case.
So, direct detection will be very difficult, but indirect detection signals (such as neutrino flux from dark matter annihilations to gauge bosons) may be observable.
 
For a strict WIMP, the possibilities for avoiding direct and indirect detection are beginning to be 
constrained. Furthermore,
these possibilities tend to involve some fine tuning.  Hence,
if the dark matter is 
strictly weakly interacting, the prospects for detection or exclusion in the near future are extremely promising.

\begin{acknowledgments}
The work of T.C. was supported in part by DOE Grants \#DE-FG02-95ER-40899 and \#DE-AC03-76SF00515 and NSF CAREER Grant \#NSF-PHY-0743315.  The work of A.P.  was supported in part by NSF CAREER Grant \#NSF-PHY-0743315  and by DOE Grant \#DE-FG02-95ER40899. The work of J.K. was supported by DOE Grant \#DE-FG02-95ER40899. The work of  D.T.S. was supported by NSF grant \#NSF-PHY-0856522.
\end{acknowledgments}


\bibliography{SingletDoubletBib}

\begin{thebibliography}{41}
\expandafter\ifx\csname natexlab\endcsname\relax\def\natexlab#1{#1}\fi
\expandafter\ifx\csname bibnamefont\endcsname\relax
  \def\bibnamefont#1{#1}\fi
\expandafter\ifx\csname bibfnamefont\endcsname\relax
  \def\bibfnamefont#1{#1}\fi
\expandafter\ifx\csname citenamefont\endcsname\relax
  \def\citenamefont#1{#1}\fi
\expandafter\ifx\csname url\endcsname\relax
  \def\url#1{\texttt{#1}}\fi
\expandafter\ifx\csname urlprefix\endcsname\relax\def\urlprefix{URL }\fi
\providecommand{\bibinfo}[2]{#2}
\providecommand{\eprint}[2][]{\url{#2}}

\bibitem[{\citenamefont{Arkani-Hamed et~al.}(2006)\citenamefont{Arkani-Hamed,
  Delgado, and Giudice}}]{ArkaniHamed:2006mb}
\bibinfo{author}{\bibfnamefont{N.}~\bibnamefont{Arkani-Hamed}},
  \bibinfo{author}{\bibfnamefont{A.}~\bibnamefont{Delgado}}, \bibnamefont{and}
  \bibinfo{author}{\bibfnamefont{G.~F.} \bibnamefont{Giudice}},
  \bibinfo{journal}{Nucl. Phys.} \textbf{\bibinfo{volume}{B741}},
  \bibinfo{pages}{108} (\bibinfo{year}{2006}), \eprint{hep-ph/0601041}.

\bibitem[{\citenamefont{Cirelli et~al.}(2006)\citenamefont{Cirelli, Fornengo,
  and Strumia}}]{Cirelli:2005uq}
\bibinfo{author}{\bibfnamefont{M.}~\bibnamefont{Cirelli}},
  \bibinfo{author}{\bibfnamefont{N.}~\bibnamefont{Fornengo}}, \bibnamefont{and}
  \bibinfo{author}{\bibfnamefont{A.}~\bibnamefont{Strumia}},
  \bibinfo{journal}{Nucl.Phys.} \textbf{\bibinfo{volume}{B753}},
  \bibinfo{pages}{178} (\bibinfo{year}{2006}), \eprint{hep-ph/0512090}.

\bibitem[{\citenamefont{Arkani-Hamed et~al.}(2005)\citenamefont{Arkani-Hamed,
  Dimopoulos, and Kachru}}]{ArkaniHamed:2005yv}
\bibinfo{author}{\bibfnamefont{N.}~\bibnamefont{Arkani-Hamed}},
  \bibinfo{author}{\bibfnamefont{S.}~\bibnamefont{Dimopoulos}},
  \bibnamefont{and} \bibinfo{author}{\bibfnamefont{S.}~\bibnamefont{Kachru}}
  (\bibinfo{year}{2005}), \eprint{hep-th/0501082}.

\bibitem[{\citenamefont{Mahbubani and Senatore}(2006)}]{Mahbubani:2005pt}
\bibinfo{author}{\bibfnamefont{R.}~\bibnamefont{Mahbubani}} \bibnamefont{and}
  \bibinfo{author}{\bibfnamefont{L.}~\bibnamefont{Senatore}},
  \bibinfo{journal}{Phys.Rev.} \textbf{\bibinfo{volume}{D73}},
  \bibinfo{pages}{043510} (\bibinfo{year}{2006}), \eprint{hep-ph/0510064}.

\bibitem[{\citenamefont{D'Eramo}(2007)}]{D'Eramo:2007ga}
\bibinfo{author}{\bibfnamefont{F.}~\bibnamefont{D'Eramo}},
  \bibinfo{journal}{Phys.Rev.} \textbf{\bibinfo{volume}{D76}},
  \bibinfo{pages}{083522} (\bibinfo{year}{2007}), \eprint{0705.4493}.

\bibitem[{\citenamefont{Enberg et~al.}(2007)\citenamefont{Enberg, Fox, Hall,
  Papaioannou, and Papucci}}]{Enberg:2007rp}
\bibinfo{author}{\bibfnamefont{R.}~\bibnamefont{Enberg}},
  \bibinfo{author}{\bibfnamefont{P.}~\bibnamefont{Fox}},
  \bibinfo{author}{\bibfnamefont{L.}~\bibnamefont{Hall}},
  \bibinfo{author}{\bibfnamefont{A.}~\bibnamefont{Papaioannou}},
  \bibnamefont{and} \bibinfo{author}{\bibfnamefont{M.}~\bibnamefont{Papucci}},
  \bibinfo{journal}{JHEP} \textbf{\bibinfo{volume}{0711}}, \bibinfo{pages}{014}
  (\bibinfo{year}{2007}), \eprint{0706.0918}.

\bibitem[{\citenamefont{Kadastik et~al.}(2010)\citenamefont{Kadastik, Kannike,
  and Raidal}}]{Kadastik:2009dj}
\bibinfo{author}{\bibfnamefont{M.}~\bibnamefont{Kadastik}},
  \bibinfo{author}{\bibfnamefont{K.}~\bibnamefont{Kannike}}, \bibnamefont{and}
  \bibinfo{author}{\bibfnamefont{M.}~\bibnamefont{Raidal}},
  \bibinfo{journal}{Phys. Rev.} \textbf{\bibinfo{volume}{D81}},
  \bibinfo{pages}{015002} (\bibinfo{year}{2010}), \eprint{0903.2475}.

\bibitem[{\citenamefont{Kadastik et~al.}(2009)\citenamefont{Kadastik, Kannike,
  and Raidal}}]{Kadastik:2009cu}
\bibinfo{author}{\bibfnamefont{M.}~\bibnamefont{Kadastik}},
  \bibinfo{author}{\bibfnamefont{K.}~\bibnamefont{Kannike}}, \bibnamefont{and}
  \bibinfo{author}{\bibfnamefont{M.}~\bibnamefont{Raidal}},
  \bibinfo{journal}{Phys. Rev.} \textbf{\bibinfo{volume}{D80}},
  \bibinfo{pages}{085020} (\bibinfo{year}{2009}), \eprint{0907.1894}.

\bibitem[{lep(2011{\natexlab{a}})}]{leptonphotonATLAS2011}
\bibinfo{journal}{ATLAS Collaboration, ATLAS-CONF-2011-13}
  (\bibinfo{year}{2011}{\natexlab{a}}).

\bibitem[{lep(2011{\natexlab{b}})}]{leptonphotonCMS2011}
\bibinfo{journal}{CMS Collaboration, CMS PAS HIG-11-02}
  (\bibinfo{year}{2011}{\natexlab{b}}).

\bibitem[{\citenamefont{Fayet}(1974)}]{Fayet:1974fj}
\bibinfo{author}{\bibfnamefont{P.}~\bibnamefont{Fayet}},
  \bibinfo{journal}{Nucl.Phys.} \textbf{\bibinfo{volume}{B78}},
  \bibinfo{pages}{14} (\bibinfo{year}{1974}).

\bibitem[{\citenamefont{Fayet}(1975)}]{Fayet:1974pd}
\bibinfo{author}{\bibfnamefont{P.}~\bibnamefont{Fayet}},
  \bibinfo{journal}{Nucl.Phys.} \textbf{\bibinfo{volume}{B90}},
  \bibinfo{pages}{104} (\bibinfo{year}{1975}).

\bibitem[{\citenamefont{Griest and Seckel}(1991)}]{Griest:1990kh}
\bibinfo{author}{\bibfnamefont{K.}~\bibnamefont{Griest}} \bibnamefont{and}
  \bibinfo{author}{\bibfnamefont{D.}~\bibnamefont{Seckel}},
  \bibinfo{journal}{Phys.Rev.} \textbf{\bibinfo{volume}{D43}},
  \bibinfo{pages}{3191} (\bibinfo{year}{1991}).

\bibitem[{\citenamefont{Aprile et~al.}(2011)}]{Aprile:2011hi}
\bibinfo{author}{\bibfnamefont{E.}~\bibnamefont{Aprile}} \bibnamefont{et~al.}
  (\bibinfo{collaboration}{XENON100 Collaboration}),
  \bibinfo{journal}{Phys.Rev.Lett.}  (\bibinfo{year}{2011}),
  \eprint{1104.2549}.

\bibitem[{\citenamefont{Felizardo et~al.}(2011)\citenamefont{Felizardo, Girard,
  Morlat, Fernandes, Giuliani et~al.}}]{Felizardo:2011uw}
\bibinfo{author}{\bibfnamefont{M.}~\bibnamefont{Felizardo}},
  \bibinfo{author}{\bibfnamefont{T.}~\bibnamefont{Girard}},
  \bibinfo{author}{\bibfnamefont{T.}~\bibnamefont{Morlat}},
  \bibinfo{author}{\bibfnamefont{A.}~\bibnamefont{Fernandes}},
  \bibinfo{author}{\bibfnamefont{F.}~\bibnamefont{Giuliani}},
  \bibnamefont{et~al.} (\bibinfo{year}{2011}), \eprint{1106.3014}.

\bibitem[{\citenamefont{Bernabei et~al.}(2008)}]{Bernabei:2008yi}
\bibinfo{author}{\bibfnamefont{R.}~\bibnamefont{Bernabei}} \bibnamefont{et~al.}
  (\bibinfo{collaboration}{DAMA}), \bibinfo{journal}{Eur. Phys. J.}
  \textbf{\bibinfo{volume}{C56}}, \bibinfo{pages}{333} (\bibinfo{year}{2008}),
  \eprint{0804.2741}.

\bibitem[{\citenamefont{Aalseth et~al.}(2011)}]{Aalseth:2011wp}
\bibinfo{author}{\bibfnamefont{C.~E.} \bibnamefont{Aalseth}}
  \bibnamefont{et~al.} (\bibinfo{year}{2011}), \eprint{1106.0650}.

\bibitem[{\citenamefont{Angloher et~al.}(2011)}]{Angloher:2011uu}
\bibinfo{author}{\bibfnamefont{G.}~\bibnamefont{Angloher}} \bibnamefont{et~al.}
  (\bibinfo{year}{2011}), \eprint{1109.0702}.

\bibitem[{\citenamefont{Ahmed et~al.}(2011)}]{Ahmed:2011gh}
\bibinfo{author}{\bibfnamefont{Z.}~\bibnamefont{Ahmed}} \bibnamefont{et~al.}
  (\bibinfo{collaboration}{CDMS Collaboration, EDELWEISS Collaboration}),
  \bibinfo{journal}{Phys.Rev.} \textbf{\bibinfo{volume}{D84}},
  \bibinfo{pages}{011102} (\bibinfo{year}{2011}), \eprint{1105.3377}.

\bibitem[{\citenamefont{Belanger et~al.}(2011)\citenamefont{Belanger, Boudjema,
  Brun, Pukhov, Rosier-Lees et~al.}}]{Belanger:2010gh}
\bibinfo{author}{\bibfnamefont{G.}~\bibnamefont{Belanger}},
  \bibinfo{author}{\bibfnamefont{F.}~\bibnamefont{Boudjema}},
  \bibinfo{author}{\bibfnamefont{P.}~\bibnamefont{Brun}},
  \bibinfo{author}{\bibfnamefont{A.}~\bibnamefont{Pukhov}},
  \bibinfo{author}{\bibfnamefont{S.}~\bibnamefont{Rosier-Lees}},
  \bibnamefont{et~al.}, \bibinfo{journal}{Comput.Phys.Commun.}
  \textbf{\bibinfo{volume}{182}}, \bibinfo{pages}{842} (\bibinfo{year}{2011}),
  \eprint{1004.1092}.

\bibitem[{\citenamefont{Giedt et~al.}(2009)\citenamefont{Giedt, Thomas, and
  Young}}]{Giedt:2009mr}
\bibinfo{author}{\bibfnamefont{J.}~\bibnamefont{Giedt}},
  \bibinfo{author}{\bibfnamefont{A.~W.} \bibnamefont{Thomas}},
  \bibnamefont{and} \bibinfo{author}{\bibfnamefont{R.~D.} \bibnamefont{Young}},
  \bibinfo{journal}{Phys.Rev.Lett.} \textbf{\bibinfo{volume}{103}},
  \bibinfo{pages}{201802} (\bibinfo{year}{2009}), \eprint{0907.4177}.

\bibitem[{\citenamefont{Essig}(2008)}]{Essig:2007az}
\bibinfo{author}{\bibfnamefont{R.}~\bibnamefont{Essig}},
  \bibinfo{journal}{Phys.Rev.} \textbf{\bibinfo{volume}{D78}},
  \bibinfo{pages}{015004} (\bibinfo{year}{2008}), \eprint{0710.1668}.

\bibitem[{\citenamefont{Hisano et~al.}(2011)\citenamefont{Hisano, Ishiwata,
  Nagata, and Takesako}}]{Hisano:2011cs}
\bibinfo{author}{\bibfnamefont{J.}~\bibnamefont{Hisano}},
  \bibinfo{author}{\bibfnamefont{K.}~\bibnamefont{Ishiwata}},
  \bibinfo{author}{\bibfnamefont{N.}~\bibnamefont{Nagata}}, \bibnamefont{and}
  \bibinfo{author}{\bibfnamefont{T.}~\bibnamefont{Takesako}},
  \bibinfo{journal}{JHEP} \textbf{\bibinfo{volume}{1107}}, \bibinfo{pages}{005}
  (\bibinfo{year}{2011}), \eprint{1104.0228}.

\bibitem[{\citenamefont{Yaguna}(2010)}]{Yaguna:2010hn}
\bibinfo{author}{\bibfnamefont{C.~E.} \bibnamefont{Yaguna}},
  \bibinfo{journal}{Phys.Rev.} \textbf{\bibinfo{volume}{D81}},
  \bibinfo{pages}{075024} (\bibinfo{year}{2010}), \eprint{1003.2730}.

\bibitem[{\citenamefont{Jarosik et~al.}(2011)}]{Jarosik:2010iu}
\bibinfo{author}{\bibfnamefont{N.}~\bibnamefont{Jarosik}} \bibnamefont{et~al.},
  \bibinfo{journal}{Astrophys. J. Suppl.} \textbf{\bibinfo{volume}{192}},
  \bibinfo{pages}{14} (\bibinfo{year}{2011}), \eprint{1001.4744}.

\bibitem[{\citenamefont{Nakamura et~al.}(2010)}]{Nakamura:2010zzi}
\bibinfo{author}{\bibfnamefont{K.}~\bibnamefont{Nakamura}} \bibnamefont{et~al.}
  (\bibinfo{collaboration}{Particle Data Group}), \bibinfo{journal}{J.Phys.G}
  \textbf{\bibinfo{volume}{G37}}, \bibinfo{pages}{075021}
  (\bibinfo{year}{2010}).

\bibitem[{\citenamefont{Desai et~al.}(2004)}]{Desai:2004pq}
\bibinfo{author}{\bibfnamefont{S.}~\bibnamefont{Desai}} \bibnamefont{et~al.}
  (\bibinfo{collaboration}{Super-Kamiokande Collaboration}),
  \bibinfo{journal}{Phys.Rev.} \textbf{\bibinfo{volume}{D70}},
  \bibinfo{pages}{083523} (\bibinfo{year}{2004}), \eprint{hep-ex/0404025}.

\bibitem[{\citenamefont{Abbasi et~al.}(2009)}]{Abbasi:2009uz}
\bibinfo{author}{\bibfnamefont{R.}~\bibnamefont{Abbasi}} \bibnamefont{et~al.}
  (\bibinfo{collaboration}{ICECUBE Collaboration}),
  \bibinfo{journal}{Phys.Rev.Lett.} \textbf{\bibinfo{volume}{102}},
  \bibinfo{pages}{201302} (\bibinfo{year}{2009}), \eprint{0902.2460}.

\bibitem[{\citenamefont{Salucci et~al.}(2010)\citenamefont{Salucci, Nesti,
  Gentile, and Martins}}]{Salucci:2010qr}
\bibinfo{author}{\bibfnamefont{P.}~\bibnamefont{Salucci}},
  \bibinfo{author}{\bibfnamefont{F.}~\bibnamefont{Nesti}},
  \bibinfo{author}{\bibfnamefont{G.}~\bibnamefont{Gentile}}, \bibnamefont{and}
  \bibinfo{author}{\bibfnamefont{C.}~\bibnamefont{Martins}},
  \bibinfo{journal}{Astron.Astrophys.} \textbf{\bibinfo{volume}{523}},
  \bibinfo{pages}{A83} (\bibinfo{year}{2010}), \eprint{1003.3101}.

\bibitem[{\citenamefont{Aprile et~al.}(2009)\citenamefont{Aprile, Baudis, Choi,
  Giboni, Lim et~al.}}]{Aprile:2008rc}
\bibinfo{author}{\bibfnamefont{E.}~\bibnamefont{Aprile}},
  \bibinfo{author}{\bibfnamefont{L.}~\bibnamefont{Baudis}},
  \bibinfo{author}{\bibfnamefont{B.}~\bibnamefont{Choi}},
  \bibinfo{author}{\bibfnamefont{K.}~\bibnamefont{Giboni}},
  \bibinfo{author}{\bibfnamefont{K.}~\bibnamefont{Lim}}, \bibnamefont{et~al.},
  \bibinfo{journal}{Phys.Rev.} \textbf{\bibinfo{volume}{C79}},
  \bibinfo{pages}{045807} (\bibinfo{year}{2009}), \eprint{0810.0274}.

\bibitem[{\citenamefont{Behnke et~al.}(2008)}]{Behnke:2008zza}
\bibinfo{author}{\bibfnamefont{E.}~\bibnamefont{Behnke}} \bibnamefont{et~al.}
  (\bibinfo{collaboration}{COUPP}), \bibinfo{journal}{Science}
  \textbf{\bibinfo{volume}{319}}, \bibinfo{pages}{933} (\bibinfo{year}{2008}),
  \eprint{0804.2886}.

\bibitem[{\citenamefont{Rajaraman et~al.}(2011)\citenamefont{Rajaraman,
  Shepherd, Tait, and Wijangco}}]{Rajaraman:2011wf}
\bibinfo{author}{\bibfnamefont{A.}~\bibnamefont{Rajaraman}},
  \bibinfo{author}{\bibfnamefont{W.}~\bibnamefont{Shepherd}},
  \bibinfo{author}{\bibfnamefont{T.~M.~P.} \bibnamefont{Tait}},
  \bibnamefont{and} \bibinfo{author}{\bibfnamefont{A.~M.}
  \bibnamefont{Wijangco}} (\bibinfo{year}{2011}), \eprint{1108.1196}.

\bibitem[{\citenamefont{Wiebusch}(2009)}]{Wiebusch:2009jf}
\bibinfo{author}{\bibfnamefont{C.}~\bibnamefont{Wiebusch}},
  \bibinfo{journal}{for the IceCube Collaboration,}  (\bibinfo{year}{2009}),
  \eprint{0907.2263}.

\bibitem[{\citenamefont{Barger et~al.}(2011)\citenamefont{Barger, Gao, and
  Marfatia}}]{Barger:2011em}
\bibinfo{author}{\bibfnamefont{V.}~\bibnamefont{Barger}},
  \bibinfo{author}{\bibfnamefont{Y.}~\bibnamefont{Gao}}, \bibnamefont{and}
  \bibinfo{author}{\bibfnamefont{D.}~\bibnamefont{Marfatia}},
  \bibinfo{journal}{Phys.Rev.} \textbf{\bibinfo{volume}{D83}},
  \bibinfo{pages}{055012} (\bibinfo{year}{2011}), \eprint{1101.4410}.

\bibitem[{\citenamefont{Gaisser}(2011)}]{DeepCoreTalk}
\bibinfo{author}{\bibfnamefont{T.}~\bibnamefont{Gaisser}},
  \bibinfo{journal}{Talk at NuTel 2011,
  \url{http://agenda.infn.it/contributionDisplay.py?contribId=53&sessionId=15&confId=3101}}
   (\bibinfo{year}{2011}).

\bibitem[{\citenamefont{Silveira and Zee}(1985)}]{Silveira:1985rk}
\bibinfo{author}{\bibfnamefont{V.}~\bibnamefont{Silveira}} \bibnamefont{and}
  \bibinfo{author}{\bibfnamefont{A.}~\bibnamefont{Zee}},
  \bibinfo{journal}{Phys. Lett.} \textbf{\bibinfo{volume}{B161}},
  \bibinfo{pages}{136} (\bibinfo{year}{1985}).

\bibitem[{\citenamefont{McDonald}(1994)}]{McDonald:1993ex}
\bibinfo{author}{\bibfnamefont{J.}~\bibnamefont{McDonald}},
  \bibinfo{journal}{Phys.Rev.} \textbf{\bibinfo{volume}{D50}},
  \bibinfo{pages}{3637} (\bibinfo{year}{1994}), \eprint{hep-ph/0702143}.

\bibitem[{\citenamefont{Burgess et~al.}(2001)\citenamefont{Burgess, Pospelov,
  and ter Veldhuis}}]{Burgess:2000yq}
\bibinfo{author}{\bibfnamefont{C.}~\bibnamefont{Burgess}},
  \bibinfo{author}{\bibfnamefont{M.}~\bibnamefont{Pospelov}}, \bibnamefont{and}
  \bibinfo{author}{\bibfnamefont{T.}~\bibnamefont{ter Veldhuis}},
  \bibinfo{journal}{Nucl.Phys.} \textbf{\bibinfo{volume}{B619}},
  \bibinfo{pages}{709} (\bibinfo{year}{2001}), \eprint{hep-ph/0011335}.

\bibitem[{\citenamefont{He and Tandean}(2011)}]{He:2011de}
\bibinfo{author}{\bibfnamefont{X.-G.} \bibnamefont{He}} \bibnamefont{and}
  \bibinfo{author}{\bibfnamefont{J.}~\bibnamefont{Tandean}}
  (\bibinfo{year}{2011}), \eprint{1109.1277}.

\bibitem[{\citenamefont{Raidal and Strumia}(2011)}]{Raidal:2011xk}
\bibinfo{author}{\bibfnamefont{M.}~\bibnamefont{Raidal}} \bibnamefont{and}
  \bibinfo{author}{\bibfnamefont{A.}~\bibnamefont{Strumia}}
  (\bibinfo{year}{2011}), \eprint{1108.4903}.

\bibitem[{\citenamefont{Tucker-Smith and Weiner}(2001)}]{TuckerSmith:2001hy}
\bibinfo{author}{\bibfnamefont{D.}~\bibnamefont{Tucker-Smith}}
  \bibnamefont{and} \bibinfo{author}{\bibfnamefont{N.}~\bibnamefont{Weiner}},
  \bibinfo{journal}{Phys.Rev.} \textbf{\bibinfo{volume}{D64}},
  \bibinfo{pages}{043502} (\bibinfo{year}{2001}), \eprint{hep-ph/0101138}.

\end{thebibliography}
\end{document}